\documentclass[a4paper,UKenglish]{lipics-v2018}
\usepackage{microtype}
\nolinenumbers

\usepackage{tablefootnote}
\usepackage{tikz}
\usetikzlibrary{shapes}

\tikzstyle{state}=[rounded rectangle,draw=none,minimum size=1.4em,inner sep=1ex]
\tikzstyle{pstate}=[rounded rectangle,draw,minimum size=1.4em,inner sep=1ex]
\tikzstyle{tran}=[draw,thick,->,>=stealth,rounded corners]
\tikzstyle{stage}=[draw,thick,rectangle,rounded corners,inner sep=1ex]

\newcommand{\calK}{\mathcal{K}}
\newcommand{\calH}{\mathcal{H}}
\newcommand{\calL}{\mathcal{L}}
\newcommand{\calM}{\mathcal{M}}
\newcommand{\calN}{\mathcal{N}}
\newcommand{\calP}{\mathcal{P}}
\newcommand{\calS}{\mathbb{S}}
\newcommand{\calO}{\mathcal{O}}
\newcommand{\bP}{\mathbb{P}}
\newcommand{\Ap}{\mathit{AP}}
\newcommand{\Ex}{\mathbb{E}}

\newcommand{\B}{\mathcal{B}}
\newcommand{\C}{\mathcal{C}}

\newcommand{\I}{\mathcal{I}}
\newcommand{\J}{\mathcal{J}}
\newcommand{\T}{\mathcal{T}}
\newcommand{\E}{\mathcal{E}}
\newcommand{\F}{\mathcal{F}}
\newcommand{\G}{\mathcal{G}}

\newcommand{\U}{\mathcal{U}}
\newcommand{\X}{\mathcal{X}}
\newcommand{\Nset}{\mathbb{N}}

\newcommand{\tr}{\mapsto}

\newcommand{\Dom}{\mathit{Dom}}
\newcommand{\Out}{\mathit{Out}}
\newcommand{\Time}{\mathit{Steps}}
\newcommand{\Comp}{\mathit{InterComplexity}}
\newcommand{\TermComp}{\mathit{ReachTerminal}}

\newcommand{\run}{\mathit{Run}}

\newcommand{\ttt}{\texttt{tt}}
\newcommand{\fff}{\texttt{ff}}

\newcommand{\Val}{\mathit{Val}}
\newcommand{\Exp}{\mathit{Exp}}

\newcommand{\Enter}{\mathit{Enter}}
\newcommand{\Stable}{\mathit{Stable}}
\newcommand{\Succ}{\mathit{Succ}}
\newcommand{\Move}{\mathit{Move}}
\newcommand{\Stages}{\mathit{Stages}}
\newcommand{\Next}{\mathit{ReachNext}}
\newcommand{\Reach}{\mathit{Reach}}
\newcommand{\Term}{\mathit{Term}}
\newcommand{\ta}{\mathit{Act}}
\newcommand{\ms}[1]{\langle #1 \rangle}
\newcommand{\ls}{[\![}
\newcommand{\rs}{]\!]}
\newcommand{\Tran}[1]{\stackrel{#1}{\hookrightarrow}}
\newcommand{\tran}[1]{{}\mathchoice%
        {\stackrel{#1}{\rightarrow}}
        {\mathop {\smash\rightarrow}\limits^{\vrule width 0pt height 0pt
                        depth 4pt\smash{#1}}}
        {\stackrel{#1}{\rightarrow}}
        {\stackrel{#1}{\rightarrow}}
        {}}

\usepackage{longtable}
\usepackage{tabu}
\usepackage[linesnumbered,ruled]{algorithm2e}
\usepackage{apxproof}

\theoremstyle{plain}
\newtheoremrep{theorem}{Theorem}
\newtheoremrep{lemma}[theorem]{Lemma}

\title{Automatic Analysis of Expected Termination Time for Population Protocols}

\author{Michael Blondin}{TU Munich, Germany}{blondin@in.tum.de}{https://orcid.org/0000-0003-2914-2734}{Supported by the Fonds de recherche du Qu\'{e}bec – Nature et technologies (FRQNT).}
\author{Javier Esparza}{TU Munich, Germany}{esparza@in.tum.de}{https://orcid.org/0000-0001-9862-4919}{Supported by ERC Advanced Grant (787367: PaVeS).}
\author{Anton\'{\i}n Ku\v{c}era}{Masaryk University, Brno, Czech Republic}{kucera@fi.muni.cz}{https://orcid.org/0000-0002-6602-8028}{Supported by the Czech Science Foundation, grant No.~P202/12/G061. The presented results were achieved during the author’s stay at TU M\"{u}nchen
	supported by the Friedrich Wilhelm Bessel Research Award (Alexander von Humboldt
	Foundation).
}
\authorrunning{M.~Blondin, J.~Esparza, A.~Ku\v{c}era}
\Copyright{Michael Blondin, Javier Esparza, and Anton\'{\i}n Ku\v{c}era}

\subjclass{
  ,
  ,
}


\supplement{Source code: \url{https://github.com/blondimi/pp-time-analysis}.}  

\EventEditors{}
\EventNoEds{0}
\EventLongTitle{}
\EventShortTitle{}
\EventAcronym{}
\EventYear{}
\EventDate{}
\EventLocation{}
\EventLogo{}
\SeriesVolume{}
\ArticleNo{}
\hideLIPIcs

\begin{document}

\maketitle

\begin{abstract}
  Population protocols are a formal model of sensor networks
  consisting of identical mobile devices. Two devices can interact and
  thereby change their states. Computations are infinite sequences of
  interactions in which the interacting devices are chosen uniformly
  at random.

  In well designed population protocols, for every initial
  configuration of devices, and for every computation starting at this
  configuration, all devices eventually agree on a consensus value. We
  address the problem of automatically computing a parametric bound on
  the expected time the protocol needs to reach this consensus. We
  present the first algorithm that, when successful, outputs a
  function $f(n)$ such that the expected time to consensus is bound by
  $\calO(f(n))$, where $n$ is the number of devices executing the
  protocol. We experimentally show that our algorithm terminates and
  provides good bounds for many of the protocols found in the
  literature.
\end{abstract}

%% Introduction
\section{Introduction}
\label{sec-intro}

Population protocols are a model of distributed computation in which agents with very limited computational resources randomly interact in pairs to perform computational tasks \cite{AngluinADFP04,DBLP:journals/dc/AngluinADFP06}. They have been used as an abstract model of wireless networks, chemical reactions, and gene regulatory networks, and it has been shown that they can be implemented at molecular level (see, e.g.,~\cite{DBLP:conf/infocom/PerronVV09,DBLP:journals/cacm/NavlakhaB15,DBLP:journals/dc/ChenCDS17,DBLP:journals/cacm/MichailS18}).

Population protocols compute by reaching a stable consensus in which all agents agree on a common output (typically a Boolean value). The output depends on the distribution of the initial states of the agents, called the initial \emph{configuration}, and so a protocol computes a predicate that assigns a Boolean value to each initial configuration. For example, a protocol in which all agents start in the same state computes the predicate $x \geq c$ if the agents agree to output $1$ when there are at least $c$ of them, and otherwise agree to output $0$. A protocol with two initial states computes the majority predicate $x \geq y$ if the agents agree to output $1$ exactly when the initial number of agents in the first state is greater than or equal to the initial number of agents in the second state.

In previous work, some authors have studied the automatic verification of population protocols.
Since a protocol has a finite state space for each initial configuration, model checking algorithms can be used to verify that the protocol behaves correctly for \emph{a finite number} of initial configurations. However, this technique cannot prove that the protocol is correct for \emph{every} configuration. In \cite{Esparza2016} it was shown that the problem of deciding whether a protocol computes some predicate, and the problem of deciding whether it computes a given predicate, are both decidable and at least as hard as the reachability problem for Petri nets. 

In practice, protocols should not only correctly compute a predicate, but also do it fast.
The most studied quantitative measure is the expected number of pairwise interactions needed to reach a stable consensus. The measure is defined for the stoichiometric model in which the pair of agents of the next interaction are picked uniformly at random. A derived measure is the \emph{parallel time}, defined as the number of interactions divided by the number of agents.
The first paper on population protocols already showed that every predicate can be computed by a protocol with expected total number of interactions $\calO(n^2 \log n)$, where $n$ is the number of agents~\cite{AngluinADFP04,DBLP:journals/dc/AngluinADFP06}. Since then, there has been considerable interest in obtaining upper and lower bounds on the number of interactions for some fundamental tasks, like \emph{leader election} and \emph{majority}, and there is also much  work on finding trade-offs between the speed of a protocol and its number of states (see, e.g., \cite{DBLP:conf/wdag/DotyS15, DBLP:conf/soda/AlistarhAEGR17,DBLP:conf/icalp/BellevilleDS17} and the references therein). However, none of these works addresses the verification~\cite{DBLP:conf/cav/ChatterjeeFM17} problem: given a protocol, determine its expected number of interactions. 

As in the qualitative case, probabilistic model checkers can be used to compute the expected number of interactions  for a given configuration. Indeed, in this case the behaviour of the protocol is captured by a finite-state Markov chain, and the expected number of interactions can be computed as the expected number of steps until a bottom strongly connected component of the chain is reached. This was the path followed in  \cite{guidelines}, using the \textsc{PRISM} probabilistic model checker. However, as in the functional case, this technique cannot give a bound valid for \emph{every} configuration. 

This paper presents the first algorithm for the automatic computation of an upper bound on the expected number of interactions. The algorithm takes advantage of the hierarchical structure of population protocols where an initial configuration reaches a stable consensus by passing through finitely many ``stages''. Entering a next stage corresponds to entering a configuration where some behavioral restrictions become permanent (for example, some interactions become permanently disabled, certain states will never be populated again, etc.). The algorithm automatically identifies such stages and computes a finite acyclic \emph{stage graph} representing the protocol evolution. If all bottom stages of the graph correspond to stabilized configurations, the algorithm proceeds by deriving bounds for the expected number of interactions to move from one stage to the next, and computes a bound for the expected number of interactions by taking an ``asymptotic maximum'' of these bounds. In unsuitable cases, the resulting upper bound can be higher than the actual expected number of interactions. We report on an implementation of the algorithm and its application to case studies.

\subparagraph*{Related work.} To the best of our knowledge, we present the first algorithm for the automatic quantitative verification of population protocols. In fact, even for sequential randomized programs, the automatic computation of the expected time is little studied. After the seminal work of Flajolet~\textit{et al}.\ in~\cite{DBLP:journals/tcs/FlajoletSZ91}, there is recent work by Kaminski~\textit{et al}.~\cite{DBLP:conf/esop/KaminskiKMO16} on the computation of expected runtimes using weakest preconditions, by Chatterje~\textit{et al}.\ on the automated analysis of recurrence relations for expected time~\cite{DBLP:conf/cav/ChatterjeeFM17}, by Van Chan Ngo \textit{et al}.~\cite{VanCH18} on the automated computation of bounded expectations using amortized resource analysis, and by Batz \textit{et al}.~\cite{DBLP:conf/esop/BatzKKM18,VanCH18} on the computation of sampling times for Bayesian networks. These works are either not targeted to distributed systems like population protocols, or do not provide the same degree of automation as ours.

\subparagraph*{Structure of the paper.} In Section~\ref{sec-prelim}, we
introduce population protocols and a simple modal logic to reason
about their behaviours. In Section~\ref{sec-evolution}, we introduce
stage graphs and explain how they allow to prove upper bounds on the
expected number of interactions of population protocols. We then give
a dedicated algorithm for the computation of stage graphs in
Section~\ref{sec-stage-compute}, analyze the bounds derived by this
algorithm in Section~\ref{sec-expected}, and report on experimental
results in Section~\ref{sec-experiments}. Finally, we conclude in
Section~\ref{sec-conclusion}.

%% Preliminaries
\section{Population protocols}
\label{sec-prelim}

In this section, we introduce population protocols and their semantics. We assume familiarity with basic notions of probability theory, such as probability space, random variables, expected value, etc. When we say that some event happens \emph{almost surely}, we mean that the probability of the event is equal to one. We use $\Nset$ to denote the set of non-negative integers. 

A \emph{population} consists of $n$ agents with states from a finite set $Q = \{A,B,\ldots\}$ interacting according to a directed \emph{interaction graph} $\G$ (without self-loops) over the agents. The interaction proceeds in a sequence of steps, where in each step an edge of the interaction graph is selected uniformly at random, and the states $(A,B)$ of the two chosen agents are updated according to a transition function containing rules of the form $(C,D) \tr (E,F)$. We assume that for each pair of states $(C,D)$, there is at least one rule $(C,D) \tr (E,F)$. If there are several rules with the same left-hand side, then one is selected uniformly at random. The unique agent identifiers are not known to the agents and not used by the protocol. 

Usually, $\G$ is considered as a \emph{complete} graph, and this assumption is adopted also in this work. Since the agent identifiers are hidden and $\G$ is complete, a population is fully determined by the number of agents in each state. Formally, a \emph{configuration} is a vector $\C \in \Nset^Q$, where $\C(A)$ is the number of agents in state~$A$. For every $A \in Q$, we use $\mathbf{1}_{A}$ to denote the vector satisfying $\mathbf{1}_{A}(A) =1$ and $\mathbf{1}_{A}(B) =0$ for all $B \neq A$.  Note that there is no difference between transitions $(A,B) \tr (C,D)$ and $(A,B) \tr (D,C)$, because both of them update a given configuration in the same way. 

Most of the population protocols studied for complete interaction graphs have a symmetric transition function where pairs $(A,B)$ and $(B,A)$ are updated in the same way. For the sake of simplicity, we restrict our attention to symmetric protocols.\footnote{All of the presented results can easily be extended to non-symmetric population protocols. The only technical difference is the way of evaluating/estimating the probability of executing a given transition in a given configuration.} Then, the transitions can be written simply as $AB\tr CD$, because the ordering of states before/after the $\tr$ symbol is irrelevant. Formally, $AB$ and $CD$ are understood as elements of $Q^{\ms{2}}$, i.e., multisets over $Q$ with precisely two elements.

\begin{definition}
	A \emph{population protocol} is a tuple $\calP = (Q,T,\Sigma,I,O)$ where
	\begin{itemize}
		\item $Q$ is a non-empty finite set of \emph{states};
		\item $T : Q^{\ms{2}} \times Q^{\ms{2}}$ is a total \emph{transition relation}; 
		\item $\Sigma$ is a non-empty finite \emph{input alphabet},
		\item $I : \Sigma \rightarrow Q$ is the \emph{input function} mapping input symbols to states,
		\item $O : Q \rightarrow \{0,1\}$ is the \emph{output function}.
	\end{itemize}
\end{definition}
We write $AB \tr CD$ to indicate that $(AB,CD) \in T$. When defining the set $T$, we usually specify the outgoing transitions only for some subset of $Q^{\ms{2}}$. For the other pairs $AB$, there (implicitly) exists a single \emph{idle} transition $AB \tr AB$. We also write $I(\Sigma)$ to denote the set $\{q \in Q \mid q = I(\sigma) \mbox{ for some } \sigma \in \Sigma\}$.

\subsection{Executing population protocols}

A transition $AB \tr CD$ is \emph{enabled} in a configuration $\C$ if $\C - \mathbf{1}_{A} - \mathbf{1}_{B} \geq \mathbf{0}$.  A transition $AB \tr CD$ enabled in $\C$ can \emph{fire} and thus produce a configuration $\C' = \C - \mathbf{1}_{A} - \mathbf{1}_{B} + \mathbf{1}_{C} + \mathbf{1}_{D}$. The \emph{probability} of executing a transition $AB \tr CD$ enabled in $\C$ is defined by 
\[
\bP[\C, AB \tr CD] = \begin{cases}
                        \frac{\C(A)\cdot(\C(A) -1)}{(n^2 -n)\cdot |\{EF \in Q^{\ms{2}} : AA \tr EF\}| } & \mbox{if } A=B\,, \\[2ex]
                        \frac{2 \cdot \C(A) \cdot \C(B)}{(n^2 -n) \cdot |\{EF \in Q^{\ms{2}} : AB \tr EF\}|} &
                        \mbox{if } A \neq B\,. 
                     \end{cases}
\]
where $n$ is the size of $\C$. Note that $2 \cdot \C(A) \cdot \C(B)$ is the number of directed edges connecting agents in states $A$ and $B$ (when $A \neq B$), and $n^2-n$ is the total number of directed edges in a complete directed graph without self-loops with $n$ vertices. If a pair of agents in states $A$ and $B$ is selected, one of the outgoing transitions of $AB$ is chosen uniformly at random.

We write $\C \tran{} \C'$ to indicate that $\C'$ is obtained from $\C$ by firing some transition, and we use $\bP[\C \tran{} \C']$ to denote the probability of executing a transition enabled in $\C$ producing $\C'$. Note that there can be several transitions enabled in $\C$ producing $\C'$, and 
$\bP[\C \tran{} \C']$ is the total probability of executing some of them.
 
An \emph{execution} initiated in a given configuration $\C$ is a finite sequence $\C_0,\ldots,\C_\ell$ of configurations such that $\ell \in \Nset$, $C_0 = \C$, and $\C_i \tran{} \C_{i+1}$ for all~$i<\ell$. A configuration $\C'$ is \emph{reachable} from a configuration $\C$ if there is an execution initiated in $\C$ ending in $\C'$. A \emph{run} is an infinite sequence of configurations $\omega = \C_0,\C_1,\ldots$ such that every finite prefix of $\omega$ is an execution. The configuration $\C_i$ of a run $\omega$ is also denoted by $\omega_i$. For a given execution $\C_0,\ldots,\C_\ell$, we use $\run(\C_0,\ldots,\C_\ell)$ to denote the set of all runs starting with $\C_0,\ldots,\C_\ell$.

For every configuration $\C$, we define the probability space $(\run(\C),\F,\bP_{\C})$, where $\F$ is the $\sigma$-algebra generated by all $\run(\C_0,\ldots,\C_\ell)$ such that $\C_0,\ldots,\C_\ell$ is an execution initiated in $\C$, and $\bP_\C$ is the unique probability measure satisfying 
\(
   \bP_{\C}(\run(C_0,\ldots,C_\ell)) \ = \ \prod_{i=0}^{\ell-1} \bP[\C_i \tran{} \C_{i+1}] \, .
\)

\subsection{A simple modal logic for population protocols}
\label{sec-modal-logic}

To specify properties of configurations, we use a qualitative variant of the branching-time logic EF. 
Let $\Ap_\calP \ = \ Q \ \cup\  \{A! \mid A \in Q \mbox{ such that there is a non-idle transition } AA \tr BC \}$.  
The formulae of our qualitative logic are constructed in the following way, where $a$ ranges over $\Ap_\calP \cup\ \{\Out_0,\Out_1\}$:
\[
   \varphi ~~ ::= ~~ a ~~|~~ \neg \varphi ~~|~~ \varphi_0 \wedge \varphi_1 
                       ~~|~~ \Box\varphi ~~|~~ \Diamond\varphi.
\]
The semantics is defined inductively:
\[
\begin{array}{lcl}
	\C \models A &   ~~\mbox{iff}~~ & \C(A) > 0, \\
	\C \models A! &   ~~\mbox{iff}~~ & \C(A) = 1, \\
	\C \models \Out_0 & ~~\mbox{iff}~~ & O(A) = 0 \mbox{ for all } A\in Q 
                                         \mbox{ such that } \C(A) > 0, \\
	\C \models \Out_1 & ~~\mbox{iff}~~ & O(A) = 1 \mbox{ for all } A\in Q 
                                         \mbox{ such that } \C(A) > 0, \\
	\C \models \neg\varphi & ~~\mbox{iff}~~ & \C \not\models \varphi, \\
	\C \models \varphi_0 \wedge \varphi_1 & ~~\mbox{iff}~~ & \C \models \varphi_0 \mbox{ and } \C \models \varphi_1, \\	
	\C \models \Box\varphi & ~~\mbox{iff}~~ & \bP_{\C}(\{\omega \in \run(\C) \mid  \omega_i \models \varphi \mbox{ for all } i \in \Nset\}) = 1, \\	
	\C \models \Diamond\varphi & ~~\mbox{iff}~~ & \bP_{\C}(\{\omega \in \run(\C) \mid  \omega_i \models \varphi \mbox{ for some } i \in \Nset\}) = 1.
\end{array}
\]
Note that $\C \models \Box\varphi$ iff all configurations reachable from $\C$ satisfy $\varphi$, and $\C \models \Diamond\varphi$ iff a run initiated in $\C$ visits a configuration satisfying~$\varphi$ almost surely (i.e., with probability one). We also use $\ttt$, $\fff$, and other propositional connectives whose semantics is defined in the standard way. Furthermore, we occasionally interpret a given set of configurations $\B$ as a formula where $\C \models \B$ iff $\C \in \B$.    

For every formula $\varphi$, we define a random variable $\Time_\varphi$ assigning to every run $\C_0,\C_1,\ldots$ either the least $\ell \in \Nset$ such that $\C_\ell \models \varphi$, or $\infty$ if there is no such~$\ell$. For a given configuration~$\C$, we use $\Ex_{\C}[\Time_\varphi]$ to denote the expected value of $\Time_\varphi$ in the probability space $(\run(\C),\F,\bP_{\C})$.

\subsection{Computable predicates, interaction complexity}
Every \emph{input} $\X \in \Nset^\Sigma$ is mapped to the
configuration $\C_{\X}$ such that $$\C_{\X}(q) =
\sum_{\substack{\sigma \in \Sigma \\ I(\sigma) = q}}
\X(\sigma)\qquad \text{for every}\ q \in Q.$$
An \emph{initial} configuration is a configuration of the form $\C_{\X}$ where $\X$ is an input. A configuration $\C$ is \emph{stable} if $\C \models \Stable$, where $\Stable \equiv (\Box \Out_0) \vee (\Box \Out_1)$. We say that a protocol $\calP$ \emph{terminates} if $\C \models  \Diamond \Stable$ for every initial configuration~$\C$.  A protocol $\calP$ \emph{computes} a unary predicate $\Lambda$ on inputs if it terminates and every stable configuration $\C'$ reachable from an initial configuration $\C_\X$ satisfies $\C' \models \Out_x$, where $x$ is either $1$ or $0$ depending on whether $\X$ satisfies $\Lambda$ or not, respectively.

The \emph{interaction complexity} of $\calP$ is a function $\Comp_\calP$ assigning to every $n \geq 1$ the \emph{maximal} $\Ex_{\C}[\Time_{\Stable}]$, where $\C$ ranges over all initial configurations of size~$n$. Since several interactions may be running in parallel, the \emph{time complexity} of $\calP$ is defined as $\Comp_\calP(n)$ divided by~$n$. Hence, asymptotic bounds on interaction complexity immediately induce the corresponding bounds on time complexity.

\subsection{Running examples} A well-studied predicate for population protocols is \emph{majority}. Here, $\Sigma = \{A,B\}$, $I(A) = A$, $I(B) = B$, and the protocol computes whether there are at least as many agents in state $B$ as there are in state~$A$. As running examples, we use two different protocols for computing majority, taken from~\cite{DV10} and~\cite{Jaax18}.

\begin{example}[(majority protocol of~\cite{DV10})]
\label{exa-pop-long}
We have that $Q = \{A,B,a,b\}$, $O(A) = O(a) = 0$, $O(B) = O(b) = 1$, and the transitions are the following:
$AB \tr ab$, $Ab \tr Aa$, $Ba \tr Bb$ and $ba \tr bb$.
\end{example}	

\begin{example}[(majority protocol of \cite{Jaax18})]
\label{exa-pop-ninja4}
Here,
$Q = \{A,B,C,a,b\}$, $O(A) = O(a) = 0$, $O(B) = O(b) = O(C) = 1$, and the transitions are the following:
$AB \tr bC$, $AC \tr Aa$, $BC \tr Bb$, $Ba \tr Bb$, $Ab \tr Aa$ and $Ca \tr Cb$.
\end{example}

%% Evolution
\section{Stages of population protocols}
\label{sec-evolution}

Most of the existing population protocols are designed so that each initial configuration passes through finitely many ``stages'' before reaching a stable configuration. Entering a next stage corresponds to performing some additional non-reversible changes in the structure of configurations.
Hence, the transition relation between stages is acyclic, and each configuration in a non-terminal stage eventually enters one of the successor stages with probability one. This intuition is formalized in our next definition.

\begin{definition}
	\label{def-stage-graph}
	Let $\calP = (Q,T,\Sigma,I,O)$ be a population protocol. A \emph{stage graph} for $\calP$ is a triple $\G = (\calS,{\Tran{}},\ls \cdot \rs)$ where $\calS$ is a finite set of \emph{stages}, ${\Tran{}} \subseteq \calS \times \calS$ is an acyclic transition relation, and $\ls \cdot \rs$ is a function assigning to each $S \in \calS$ a set of configurations $\ls S \rs$ such that the following conditions are satisfied: 
	\begin{enumerate}
		\item[(a)] For every initial configuration $\C$ there is some $S \in \calS$ such that $\C \in \ls S \rs$.
		\item[(b)] For every $S \in \calS$ with at least one successor under $\Tran{}$, and for every $\C \in \ls S \rs$, we have that\footnote{Recall that sets of configurations can be interpreted as formulae of the modal logic introduced in Section~\ref{sec-modal-logic}.} 
		   $\C \models \Diamond \Succ(S)$, where $\Succ(S) \equiv \bigvee_{S \Tran{} S'} \ls S' \rs$. 
	\end{enumerate}
\end{definition}

\noindent
Note that a stage graph for $\calP$ is not determined uniquely. Even a trivial graph with one stage $S$ and no transitions such that $\ls S \rs$ is the set of all configurations is a valid stage graph by Definition~\ref{def-stage-graph}. To analyze the interaction complexity of $\calP$, we need to construct a stage graph so that the expected number of transitions needed to move from stage to stage can be determined easily, and all terminal stages consist only of stable configurations (see Lemma~\ref{lem-exp-time} below).

Formally, a stage $S$ is \emph{terminal} if it does not have any successors, i.e., there is no $S'$ satisfying $S \Tran{} S'$. Let $\T$ be the set of all terminal stages, and let $\Term \equiv \bigvee_{S \in \T} \ls S \rs$. It follows directly from Definition~\ref{def-stage-graph}(b) that $\C \models \Diamond \Term$ for every initial configuration $\C$. Let $\TermComp_\G$ be a function assigning to every $n \geq 1$ the maximal $\Ex_{\C}[\Time_{\Term}]$, where $\C$ ranges over all initial configurations of size~$n$. Furthermore, for every $S \in \calS$, we define a function $\Next_S$ assigning to every $n \geq 1$ the maximal $\Ex_{\C}[\Time_{\Succ(S)}]$, where $\C$ ranges over all configurations of $\ls S \rs$ of size~$n$ (if $\ls S \rs$ does not contain any configuration of size $n$, we put $\Next_S(n) = 0$).

An asymptotic upper bound for $\TermComp_\G$ can be obtained by
developing an asymptotic upper bound for all $\Next_S$, where $S \in
\calS$. Even though such a bound on $\TermComp_\G$ depends on
$|\calS|$, the latter is a constant since it is independent from the
number of agents. Therefore, the following holds:

\begin{lemmarep}
\label{lem-upper}
	Let $\calP = (Q,T,\Sigma,I,O)$ be a population protocol and $\G = (\calS,{\Tran{}},\ls \cdot \rs)$ a stage graph for $\calP$. Let $f : \Nset \rightarrow \Nset$ be a function such that $\Next_S \in \calO(f)$ for all $S \in \calS$. Then $\TermComp_\G \in \calO(f)$.
\end{lemmarep}
\begin{proof}
	Let $\C_0$ be an initial configuration. For every $i \in \Nset$, we define random variables $\Move_i$ and $\Stages_i$ over the runs initiated in $\C_0$ inductively as follows. Let $\omega = \C_0,\C_1,\ldots$ be a run initiated in in $\C$. Then
	\begin{itemize}
		\item $\Move_0(\omega) = 0$, $\Stages_0(\omega) = \{S \in \calS \mid \C_0 \in \ls S \rs\}$.
		\item Let $M = \{S' \in \calS \mid S \Tran{} S' \mbox{ for some } S \in \Stages_i(\omega)\}$.
		   If $M = \emptyset$, we put $\Move_{i+1}(\omega) = 0$ and $\Stages_{i+1}(\omega) = \Stages_i(\omega)$. Otherwise, let $k = \sum_{j=0}^i \Move_j(\omega)$. We define $\Move_{i+1}(\omega)$ as the least $\ell \in \Nset$ such that $\C_{k+\ell} \in \bigcup_{S \in M} \ls S \rs$, or  $\infty$ if no such $\ell \in \Nset$ exists (this includes the case when $k = \infty$). Furthermore, if $\Move_{i+1}(\omega) < \infty$, we put $\Stages_{i+1}(\omega) = \{S \in M \mid \C_{k+\Move_i} \in \ls S \rs\}$; otherwise, $\Stages_{i+1}(\omega) = \Stages_i(\omega)$.
	\end{itemize}
    Condition~(b) of Definition~\ref{def-stage-graph} immediately implies $\bP_{\C_0}[\Move_i {=} \infty] = 0$ for all $i \in \Nset$. Since $\Tran{}$ is acyclic, for almost all runs $\omega$ initiated in $\C_0$ we have that $\Move_i(\omega) = 0$ for every $i \geq |\calS|$. Thus, we obtain 
    \[
       \Ex_{\C_0}[\Time_{\Term}] \ \leq \ \Ex_{\C_0}\left[\sum_{i=0}^\infty \Move_i\right] \ = \ \Ex_{\C_0}\left[\sum_{i=0}^{|\calS|} \Move_i\right] \ = \ \sum_{i=0}^{|\calS|}\Ex_{\C_0}[\Move_i] \,.
    \]
    Clearly, $\Ex_{\C_0}[\Move_i] \leq \max_{S \in \calS} \Ex_{\C}[\Time_{\Succ(S)}]$ where $\C$ ranges over all configurations of $\ls S \rs$ whose size is equal to the size of $\C_0$. In other words,  $\Ex_{\C_0}[\Move_i] \leq \max_{S \in \calS} \Next_S(n)$, where $n$ is the size of $\C_0$.  Since $\Next_S \in \calO(f)$ for all $S \in \calS$ and $|\calS|$ is a constant, we obtain $\TermComp_\G \in \calO(f)$.
\end{proof}

Observe that if every terminal stage $S$ satisfies $\ls S \rs
\subseteq \Stable$, then $\Comp_\calP \leq \TermComp_\G$
(pointwise). Thus, we obtain the following:

\begin{lemma}
\label{lem-exp-time}
  Let $\calP = (Q,T,\Sigma,I,O)$ be a population protocol and $\G = (\calS,{\Tran{}},\ls \cdot \rs)$ a stage graph for $\calP$ such that $\ls S \rs \subseteq \Stable$ for every terminal stage $S$. Let $f : \Nset \rightarrow \Nset$ be a function such that $\Next_S \in \calO(f)$ for all $S \in \calS$. Then $\Comp_\calP \in \calO(f)$.
\end{lemma}

\begin{figure}[t]
	\centering%
	\begin{tikzpicture}[x=5.5cm,y=1.5cm,font=\footnotesize,scale=.8]
	\node[stage,label=left:{$S_0$}] (S0) at (0,0)    {$\neg a \wedge \neg b \wedge \neg C$};
	\node[stage,label=left:{$S_1$}] (S1) at (-1,-1)  {$\Box(A \wedge \neg B)$};
	\node[stage,label=left:{$S_2$}] (S2) at (0,-1)   {$\Box(\neg A \wedge B)$};
	\node[stage,label=left:{$S_3$}] (S3) at (1,-1)   {$\Box(\neg A \wedge \neg B \wedge C)$};
	\node[stage,label=left:{$S_4$}] (S4) at (-1,-2)  {$\Box(A \wedge \neg B \wedge \neg C)$};
    \node[stage,label=left:{$S_5$}] (S5) at (0,-2)   {$\Box(\neg A \wedge B \wedge \neg C)$};
    \node[stage,label=left:{$S_6$}] (S6) at (1,-2)   {$\Box(\neg A \wedge \neg B \wedge C \wedge \neg a)$};
	\node[stage,label=left:{$S_7$}] (S7) at (-1,-3)  {$\Box(A \wedge \neg B \wedge \neg C \wedge \neg b)$};
    \node[stage,label=left:{$S_8$}] (S8) at (0,-3)   {$\Box(\neg A \wedge B \wedge \neg C \wedge \neg a)$};
    \draw [tran] (S0) -- (S1);  
    \draw [tran] (S0) -- (S2);  
    \draw [tran] (S0) -- (S3);  
    \draw [tran] (S1) -- (S4);  
    \draw [tran] (S2) -- (S5);  
    \draw [tran] (S3) -- (S6);  
    \draw [tran] (S4) -- (S7);  
    \draw [tran] (S5) -- (S8);  
	\end{tikzpicture}  
	\caption{A stage graph for the majority protocol of Example~\ref{exa-pop-ninja4}.}
	\label{fig-stage-graph}	
\end{figure}
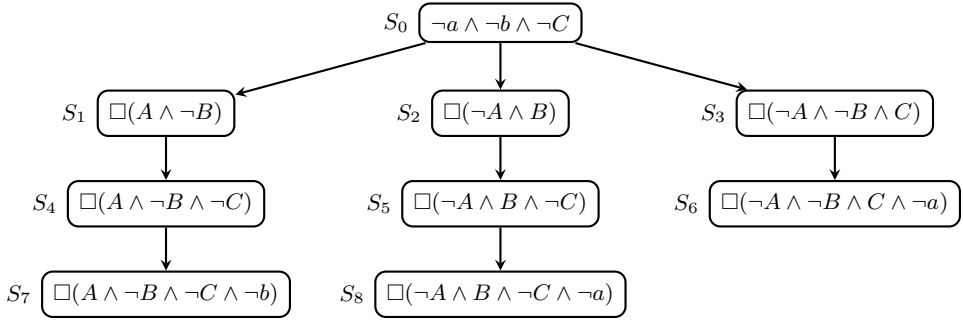

\subsection{An example of a stage graph}

In this section, we give an example of a stage graph $\G$ for the majority protocol $\calP$ of Example~\ref{exa-pop-ninja4}, and we show how to analyze the interaction complexity of $\calP$ using~$\G$. 

The stage graph $\G$ of Fig.~\ref{fig-stage-graph} is a simplified version of the stage graph computed by the algorithm of the forthcoming Section~\ref{sec-stage-compute}.  Intuitively, the hierarchy of stages corresponds to ``disabling more and more states'' along runs initiated in initial configurations. For each stage $S_i$ of $\G$, the set $\ls S_i \rs$ consists of all configurations satisfying the associated formula shown in Fig.~\ref{fig-stage-graph}. Since $\ls S_0 \rs$ is precisely the set of all initial configurations, Condition~(a) of Definition~\ref{def-stage-graph} is satisfied. For every $\C_0 \in \ls S_0 \rs$, transition $AB \tr bC$ can be executed in all configurations reachable from $\C_0$ until $A$ or $B$ disappears. Furthermore, the number of $A$'s and $B$'s can only decrease along every run initiated in $\C_0$. Hence, $\C_0$ almost surely reaches a configuration $\C$ where $A$ or $B$ (or both of them) disappear. Note that if, e.g., $\C(A)=0$ and $\C(B) >0$, then this property is ``permanent'', i.e., every successor $\C'$ of $\C$ also satisfies $\C'(A)=0$ and $\C'(B) >0$. Thus, we obtain the stages $S_1$, $S_2$, and $S_3$. Observe that if $A$ and $B$ disappear simultaneously (which happens iff the initial configuration $\C_0$ satisfies $\C_0(A) = \C_0(B)$), then the configuration $\C$ will contain at least one copy of $C$ which cannot be removed. 

In all configurations of $\ls S_1 \rs$, the only potentially executable transitions are the following: $AC \tr Aa$, $Ab \tr Aa$, $Ca \tr Cb$.
Since $A$ appears in all configurations reachable from configurations of $\ls S_1 \rs$, the transition $AC \tr Aa$ stays enabled in all of these configurations until $C$ disappears. Hence, every configuration of $\ls S_1 \rs$ almost surely reaches a configuration of  $\ls S_4 \rs$. Similarly, we can argue that all configurations of  $\ls S_4 \rs$ almost surely reach a configuration of $\ls S_7 \rs$, etc. Hence, Condition~(b) of Definition~\ref{def-stage-graph} is also satisfied.

Let $\C_0 \in \ls S_0 \rs$ be an initial configuration of size $n$, and let $\C$ be a configuration reachable from $\C_0$ such that $m = \min\{\C(A),\C(B)\} > 0$. The probability of firing $AB \tr bC$ stays larger than $m^2/n^2$ in all configurations reached from $\C$ by executing a finite sequence of transitions \emph{different} from $AB \tr bC$. This means that $AB \tr bC$ is fired after at most $n^2/m^2$ trials on average. Since $\min\{\C_0(A),\C_0(B)\} \leq n/2$, we obtain
\[
   \Next_{S_0}(n) \quad \leq \quad \sum_{i=1}^{n/2} \frac{n^2}{i^2} \quad \leq \quad n^2 \cdot \sum_{i=1}^{n} \frac{1}{i^2} \quad \leq \quad n^2 \cdot \calH_{n,2} \quad \in \quad  \calO(n^2)\,.
\]
Here, $\calH_{n,2}$ is the $n$-th Harmonic number of order~$2$. As $\lim_{n\rightarrow \infty} \calH_{n,2} = c < \infty$, we have that $n^2 \cdot \calH_{n,2} \in \calO(n^2)$.

Now, let us analyze $\Next_{S_1}(n)$. Let $\C \in \ls S_1 \rs$ be a configuration of size~$n$. We need to fire the transition $AC \tr Aa$ repeatedly until all $C$'s disappear. Let $\C'$ be a configuration reachable from $\C$ such that $\C'(C) = m$. Since $\C \models \Box(A \wedge \neg B)$, we have that $\C'(A) > 0$, and hence the probability of firing  $AC \tr Aa$ in $\C'$ is at least $m/n^2$. Thus, we obtain
\[
 \Next_{S_1}(n) \quad \leq \quad \sum_{i=1}^{n} \frac{n^2}{i} \quad \leq \quad n^2 \cdot \sum_{i=1}^{n} \frac{1}{i} \quad \leq \quad n^2 \cdot \calH_{n} \quad \in \quad  \calO(n^2 \log(n))\,.
\] 
Here $\calH_{n}$ denotes the $n$-th Harmonic number (of order $1$). Since $\lim_{n \rightarrow \infty} \calH_n = c \cdot \log(n)$ where $c$ is a constant, we get $n^2 \cdot \calH_{n} \in  \calO(n^2 \log(n))$.

Similarly, we can show that $\Next_{S_i}(n) \in \calO(n^2 \log(n))$ for every stage $S_i$ of the considered stage graph. Since all configurations associated to terminal stages are stable, we can apply Lemma~\ref{lem-exp-time} and conclude that $\Comp_\calP \in \calO(n^2 \log(n))$. Let us note that the algorithm of  the forthcoming Section~\ref{sec-stage-compute} can derive this result fully automatically in less than a second. 

\section{Computing a stage graph}
\label{sec-stage-compute}

In this section, we give an algorithm computing a stage graph for a given population protocol. Intuitively, the algorithm tries to identify a subset of transitions which will be simultaneously and permanently disabled in the future with probability one, and also performs a kind of ``case analysis'' how this can happen. The resulting stage graph admits computing an upper asymptotic bounds on $\Next_S$ for every stage $S$, which allows to compute an asymptotic upper bound on the interaction complexity of the protocol by applying Lemma~\ref{lem-exp-time}.

For the rest of this section, we fix a population protocol $\calP = (Q,T,\Sigma,I,O)$. A \emph{valuation} is a \emph{partial} function \mbox{$\nu : \Ap_\calP \rightarrow \{\ttt,\fff\}$} such that $\nu(A!) = \ttt$ implies $\nu(A) = \ttt$ whenever $A!,A \in \Dom(\nu)$, where $\Dom(\nu)$ is the domain of~$\nu$. Slightly abusing our notation, we also denote by $\nu$ the  propositional formula 
\[
    \bigwedge_{\substack{p \in \Dom(\nu)\\ \nu(p)=\ttt}} p \quad \wedge \quad 
    \bigwedge_{\substack{p \in \Dom(\nu)\\ \nu(p)=\fff}} \neg p
\]
Hence, by writing $\C \models \nu$ we mean that $\C$ satisfies the above formula.

For every \emph{transition head} $AB \in Q^{\ms{2}}$, let $\xi_{AB}$ be either the formula $\neg A \vee \neg B$ or the formula $\neg A \vee A!$, depending on whether $A \neq B$ or $A = B$, respectively. Hence, the formulae $\xi_{AB}$ and $\neg\xi_{AB}$ say that all transitions of the form $AB \tr CD$ are disabled and enabled, respectively. For a given set $\T \subseteq Q^{\ms{2}}$, consider the propositional formula $\Psi_\T \equiv \bigwedge_{AB \in \T} \xi_{AB}$. To simplify our notation, we write just $\T$ instead of $\Psi_{\T}$, i.e., $\C \models \T$ iff all transitions specified by $\T$ are disabled in~$\C$.

\begin{definition}
\label{def-stage}
    Let $\calP = (Q,T,\Sigma,I,O)$ be a population protocol.
	A \emph{$\calP$-stage} is a triple $S = (\Phi,\pi,\T)$ where 
	\begin{itemize}
		\item $\Phi$ is a propositional formula over $\Ap_\calP$, 
		\item $\pi$ is a valuation, called the
                  \emph{persistent valuation},
		\item $\T \subseteq Q^{\ms{2}}$ is a set of transition heads, called the \emph{permanently disabled transition heads}.
	\end{itemize}
    For every $\calP$-stage $S = (\Phi,\pi,\T)$, we put $\ls S \rs = \{\C \mid \C \models  \Phi \wedge \Box\pi \wedge \Box\T\}$. 
\end{definition}

Our algorithm computes a stage graph 
for $\calP$ gradually by adding more and more \mbox{$\calP$-stages}. It starts by inserting the \emph{initial} $\calP$-stage $S_0 = (\Phi,\emptyset,\emptyset)$, where 
\[
  \Phi \ \equiv \ \left(\bigvee_{A\in I(\Sigma)} A\right) \ \wedge \ \bigwedge_{A \in Q\smallsetminus I(\Sigma)} \neg A \,.
\]  
Note that $\ls S_0 \rs$ is precisely the set of all initial configurations (the empty conjunction is interpreted as \textit{true}). Then, the algorithm picks an unprocessed $\calP$-stage in the part of the stage graph constructed so far, and computes its immediate successors. This goes on until all $\calP$-stages become either internal or terminal. Since the total number of constructed $\calP$-stages can be exponential in the size of $\calP$, the worst-case complexity of our algorithm is exponential. However, as we shall see in Section~\ref{sec-experiments}, protocols with hundreds of states and transitions can be successfully analyzed even by our prototype implementation.  

Let $S = (\Phi,\pi,\T)$ be a non-terminal $\calP$-stage, and let $\Ap_{S} \subseteq \Ap_{\calP}$ be the set of all atomic propositions appearing in the formula $\Phi$. The successor $\calP$-stages of $S$ are constructed as follows. First, the algorithm computes the set $\Val_S$ consisting of all valuations $\nu$ with domain $\Ap_{S}$ such that $\nu$ satisfies $\Phi$ when the latter is interpreted over $\Ap_{S}$. Intuitively, this corresponds to dividing $\ls S \rs$ into disjoint ``subcases'' determined by different $\nu$'s (as we shall see, $\Phi$ always implies the formula $\pi \wedge \T$, so $\nu$ cannot be in conflict with the information represented by $\pi$ and $\T$; furthermore, we have $\Dom(\pi) \subseteq \Dom(\nu)$). Then, for each $\nu \in \Val_S$, a $\calP$-stage $S_\nu$ is constructed, and $S_\nu$ may or may not become a successor of $S$. If none of these $S_\nu$ becomes a successor of $S$, then $S$ is declared as terminal. 

Let us fix some $\nu \in \Val_S$. In the rest of this section, we show how to compute the $\calP$-stage $S_\nu = (\Phi_\nu,\pi_\nu,\T_\nu)$, and how to determine whether or not $S_\nu$ becomes a successor of~$S$. An explicit pseudocode for constructing $S_\nu$ is given in in the appendix.

\subsection{Computing the valuation $\pi_\nu$} The valuation $\pi_\nu$ is obtained by extending $\pi$ with the ``permanent part'' of $\nu$. Intuitively, we try to identify $A \in Q$ such that $\nu(A) = \ttt$ (or $\nu(A) = \fff$) and all transitions containing $A$ on the left-hand (or the right-hand) side are permanently disabled. Furthermore, we also try to identify $A \in Q$ such that $\nu(A!) = \ttt$ and the number of $A$'s cannot change by firing transitions which are not permanently disabled. Technically, this is achieved by a simple fixed-point computation guaranteed to terminate quickly. The details are given in the appendix.

\begin{toappendix}
	
\subsection{The procedure for computing the valuation $\pi_\nu$.}

First, we show how to compute two sets $\calM \subseteq Q$ and $\calN \subseteq Q$ satisfying the following properties: 
\begin{itemize}
\item[(1)] $\nu(A) = \fff$  for every $A \in \cal M$, and $\nu(A!) = \ttt$ for every  $A \in \calN$.\\
(Every configuration satisfying $\nu$ puts no agents in states of $\calM$, and exactly one agent in each state of $\calN$.) 
\item[(2)] For every configuration $\C \in \ls S \rs$ such that $\C \models \nu$ and for every configuration $\C'$ reachable from $\C$:   $\C' \models \neg A$ for every $A \in \cal M$, and $\C' \models A!$ for every $A \in \calN$. \\
(Every configuration reachable from a configuration satisfying $\nu$ puts no agents in states of $\calM$, and exactly one agent in each state of $\calN$.)
\end{itemize}
The pair $(\calM,\calN)$ is computed as the greatest fixed-point of a function $f \colon 2^Q \times 2^Q \rightarrow 2^Q \times 2^Q$. Intuitively, we start with the pair of sets $(M_0, N_0)$ such that
$A \in M_0$  if{}f $\nu(A) = \fff$ and $A \in N_0$ if{}f $\nu(A!) = \ttt$, i.e., with largest pair of sets satisfying (1). Then we repeatedly remove states for which we can determine that (2) does not hold. For example, if
$M_0=\{A, B\}$ and the protocol has a transition $CD \mapsto AD$, then we can remove $A$ from $M_0$, because there exists a configuration $\C$ satisfying $\neg A \wedge \neg B$, from which we can reach a configuration satisfying $A$. 

Formally, for a given pair $(M,N) \in 2^Q \times 2^Q$, let $f$ be the function that returns the pair $(M',N')$ given by:
\begin{itemize}
	\item the set $M'$ consists of all $A \in Q$ where $\nu(A) = \fff$ and every transition of the form $CD \tr AB$ satisfies either $\{C,D\} \cap M \neq \emptyset$, or $CD \in \T$, or $C=D$ and $C \in N$;
	\item the set $N'$ consists of all $A \in Q$ where $\nu(A!) = \ttt$, and the following conditions are satisfied:
	\begin{itemize}
		\item Let $AB \tr CD$ be a transition such that $A \neq B$ and $C \neq A \neq D$. Then $B \in M$ or $AB \in \T$.
		\item Let $AB \tr AA$ be a transition such that $A \neq B$. Then $B \in M$ or $AB \in \T$.
		\item Let $CD \tr AB$ be a transition such that $C \neq A \neq D$. Then either $\{C,D\} \cap M \neq \emptyset$, or $CD \in \T$, or $C=D$ and $C \in N$.
	\end{itemize}
\end{itemize}
Observe that $f$ is monotone, hence the greatest fixed-point $(\calM,\calN)$ of $f$ exists and can be computed in polynomial time. Further, let $\E$ be the set of all $A \in Q$ such that $\nu(A) =\ttt$ and every transition of the form $AB \tr CD$, where $C \neq A \neq D$, satisfies either $B \in \calM$, or $AB \in \T$, or $A = B$ and $A \in \calN$. Intuitively, this is the set of states that must necessarily contain exactly one agent, and so we put
$\pi_\nu(A) = \fff$ for all $A \in \calM$, $\pi_\nu(A) = \ttt$ for all $A \in \E$, and $\pi_\nu(A!) = \ttt$ for all $A \in \calN$.
\end{toappendix}

\begin{toappendix}
\begin{example}
\label{exa-majority-pi}
Let $\calP$ be the protocol of Example~\ref{exa-pop-long}, and let $S = (\Phi,\emptyset,\emptyset)$ be the initial $\calP$-stage, where $\Phi \equiv (A \vee B) \wedge \neg a \wedge \neg b$. There are three valuations $\nu_A,\nu_B,\nu_{AB}$ satisfying $\Phi$, which set to $\ttt$ precisely the variable $A$, or $B$, or both $A$ and $B$, respectively. The fixed-point computation starts 
from the sets $(\{B, a, b\}, \emptyset)$, $(\{A, a, b\}, \emptyset)$, and $(\{a, b\}, \emptyset)$, respectively. The greatest fixed-point $(\calM,\calN)$ is $(\{B, a, b\}, \emptyset)$, $(\{A, a, b\}, \emptyset)$, and  $(\emptyset, \emptyset)$, respectively. We have $\Dom(\pi_{\nu_A}) = \Dom(\pi_{\nu_B}) = \{A,B,a,b\}$ and $\Dom(\pi_{\nu_{AB}}) = \emptyset$.  
\end{example}
\end{toappendix}

\subsection{Computing the set $\T_\nu$ and the formula $\Phi_\nu$}

In some cases, the constructed persistent valuation $\pi_\nu$ already guarantees that a configuration satisfying $\pi_\nu \wedge \T$ is stable or cannot evolve (fire non-idle transitions) any further. Then, we in fact identified a subset of configurations belonging  to $\ls S \rs$ which does not require any further analysis. Hence, we put  $\T_\nu = \T$,  $\Phi_\nu = \pi_\nu$, and the configuration $S_\nu$ becomes a successor $\calP$-stage of $S$ declared as terminal. 

Formally, we say that $(\pi_\nu,\T)$ is \emph{stable} if there is $x \in \{0,1\}$ such that for all states $A \in Q$ where $\pi_\nu(A) = \ttt$ or $A \not\in \Dom(\pi_\nu)$ we have that $\Out(A) = x$, and for every transition $CD \tr EF$ where $\Out(E) \neq x$ or $\Out(F) \neq x$, the formula $(\pi_\nu \wedge \T) \Rightarrow \xi_{CD}$ is a propositional tautology.
Furthermore, we say that $(\pi_\nu,\T)$ is \emph{dead} if it is not stable and for every non-idle transition $CD \tr EF$ we have that the formula $(\pi_\nu \wedge \T) \Rightarrow \xi_{CD}$ is a propositional tautology. 

If $S_\nu$ is \emph{not} stable or dead, we use $\pi_\nu$ and $\T$ to compute the \emph{transformation graph} $G_\nu$, and then analyze $G_\nu$ to determine $\T_\nu$ and $\Phi_\nu$.

\subsubsection{The transformation graph}
\label{sec-trans-graph}

The vertices of the transformation graph $G_\nu$ are the states which have \emph{not} yet been permanently disabled according to $\pi_{\nu}$, and the edges are determined by a set of transitions whose heads have not yet been permanently disabled according to $\pi_\nu$ and $\T$. Formally, 
we put $G_\nu = (V,\tran{})$ where the set of vertices  $V$ consists of all $A \in Q$ such that either $A \not\in \Dom(\pi_{\nu})$ or $\pi_{\nu}(A) = \ttt$, and the set of edges is determined as follows: Let $AB \tr CD$ be a non-idle transition such that $(\pi_\nu \wedge \T) \Rightarrow \xi_{AB}$ is \emph{not} a tautology.
\begin{itemize}
	\item If the sets $\{A,B\}$ and $\{C,D\}$ are disjoint, then the transition generates the edges $A \tran{} C$,  $A \tran{} D$, $B \tran{} C$,  $B \tran{} D$. Intuitively, both $A$ and $B$ can be ``transformed'' into $C$ or $D$.
	\item Otherwise, the transition has the form $AB \tr AD$ for $B \neq D$. In this case it generates the edge $B \tran{} D$. Intuitively, $B$ can be ``transformed'' into $D$ in the context of $A$.  
\end{itemize}

\begin{figure}[t]
	\centering%
	\begin{tikzpicture}[x=1.6cm,y=1.6cm,font=\small]
	\node[state] (A) at (0,0)  {$A$};
	\node[state] (B) at (1,0)  {$B$};
	\node[state] (a) at (0,-1) {$a$};
	\node[state] (b) at (1,-1) {$b$};
	\draw [tran] (A) -- (a); 
	\draw [tran] (A) -- (b); 
	\draw [tran] (B) -- (a); 
	\draw [tran] (B) -- (b); 
	\draw [tran] (a) --  (b); 
	\draw [tran] (b) -- +(0,-.3) -- +(-1,-.3) -- (a); 
	\begin{scope}[x=2.5cm,shift={(2,0)}]
	\node[state] (A) at (0,0)  {$A$};
	\node[state] (B) at (1,0)  {$B$};
	\node[state] (C) at (0,-.5) {$C$};
	\node[state] (a) at (0,-1) {$a$};
	\node[state] (b) at (1,-1) {$b$};
	\draw [tran] (A) --  (C); 
	\draw [tran] (A) -- (b); 
	\draw [tran] (B) --  (C); 
	\draw [tran] (B) --  (b); 
	\draw [tran] (a) --  (b); 
	\draw [tran] (b) -- +(0,-.3) --  +(-1,-.3) -- (a); 
	\draw [tran] (C) --  (a); 
	\draw [tran] (C) --  (b); 
	\end{scope}
	\end{tikzpicture}  
	\caption{Transformation graphs  of Example~\ref{exa-majority-TG}.}
	\label{fig-init-stages}	
\end{figure}
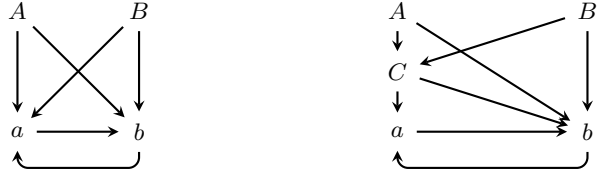

\begin{example}\label{exa-majority-TG}
Consider the protocol of Example~\ref{exa-pop-long} and its initial stage $S = (\Phi, \pi, \T)$ where $\Phi = (A \vee B) \wedge \neg a \wedge
\neg b$ and $\pi = \T = \emptyset$. Three valuations satisfy $\Phi$;
in particular the valuation $\nu$ which sets to $\ttt$ precisely the
variables $A$ and $B$. Since both $A$ and $B$ can disappear in the future, and both $a$ and $b$ can become populated, the ``permanent part'' of $\nu$, i.e., the valuation $\pi_{\nu}$, has the empty domain. The transformation graph $G_\nu$ is shown in Fig.~\ref{fig-init-stages}~(left).

Consider now the majority protocol of Example~\ref{exa-pop-ninja4} with initial stage $(\Phi,\emptyset,\emptyset)$ (where $\Phi$ says there are only $A$'s and $B$'s), and a valuation $\nu$ which  sets to $\ttt$ precisely the variables $A$ and $B$. The domain of $\pi_{\nu}$ is again the empty set, and the transformation graph $G_\nu$ is shown in Fig.~\ref{fig-init-stages}~(right).	
\end{example}

A key observation about transformation graphs is that all transitions generating edges connecting two \emph{different} strongly connected components (SCCs) of $G_\nu$ become simultaneously disabled in the future almost surely. More precisely, let $\Exp_\nu$ be the set of all $AB \in Q^{\ms{2}}$ such that there exists a transition $AB \tr CD$ generating an edge of $G_\nu$ connecting two different SCCs of $G_\nu$. We have the following:

\begin{lemma}
\label{lem-trans}
	Let $G_\nu$ be a transformation graph, and let $\C$ be a configuration such that $\C \models  \Box\pi_{\nu} \wedge \Box \T$. Then $\C \models \Diamond \Exp_\nu$. Furthermore, $\C \models \Diamond\Box \Exp_\nu$.
\end{lemma}  
However, there is a subtle problem. When the transitions specified by $\Exp_\nu$ become simultaneously disabled \emph{for the first time}, they may be disabled only \emph{temporarily}, i.e., $\C$ does \emph{not} have to satisfy the formula $\Box(\Exp_\nu \Rightarrow \Box\Exp_\nu)$. As we shall see in Section~\ref{sec-expected}, it is relatively easy to obtain an upper bound on the expected number of transitions needed to visit a configuration satisfying $\Exp_\nu$. However, it is harder to give an upper bound on the expected number of transitions needed to reach a configuration satisfying $\Box\Exp_\nu$ (i.e., entering the next stage) unless $\C \models \Box(\Exp_\nu \Rightarrow \Box \Exp_\nu)$. This difficulty is addressed in the next section.

\begin{example}
	We continue with Example~\ref{exa-majority-TG}. For the transformation graph of Fig.~\ref{fig-init-stages}~(left), we have $\Exp_{\nu} = \{AB\}$. For the transformation graph of Fig.~\ref{fig-init-stages}~(right), we have $\Exp_\nu = \{AB,AC,BC\}$. Hence, according to Lemma~\ref{lem-trans}, every initial configuration of the majority protocol of Example~\ref{exa-pop-long} almost surely reaches a configuration satisfying $\neg A \vee \neg B$, and every initial configuration of the majority protocol of Example~\ref{exa-pop-ninja4} almost surely reaches a configuration satisfying $(\neg A \vee \neg B) \wedge (\neg A \vee \neg C) \wedge (\neg B \vee \neg C)$. Furthermore, in both cases $\C \models \Box(\Exp_\nu \Rightarrow \Box \Exp_\nu)$ for every initial configuration~$\C$.
\end{example}

\subsubsection{Computing $\T_\nu$ and $\Phi_\nu$: Case $\Exp_\nu \neq \emptyset$} Let 
$\Gamma_\nu \ \equiv \ \nu \ \wedge \ \Box\pi_\nu \ \wedge\ \Box\T$, and let $\C$ be a configuration satisfying $\Gamma_\nu$. A natural idea to construct $\T_\nu$ is to enrich $\T$ by~$\Exp_\nu$. However, $\Exp_\nu$ can be empty, i.e., the transformation graph $G_\nu$ may consist just of disconnected SCCs. For this reason we first consider the case where 
$\Exp_\nu$ is nonempty.
\bigskip

\noindent
\textbf{Computing  $\T_\nu$}.\quad As discussed in Section~\ref{sec-trans-graph}, the fact that $\C \models \Diamond \Box \Exp_\nu$ does not necessarily imply $\C \models \Box(\Exp_\nu \Rightarrow \Box \Exp_\nu)$ complicates the interaction complexity analysis. Therefore, after computing $\Exp_\nu$ we try to compute a non-empty subset $\J_\nu \subseteq \Exp_\nu$ such that $\C \models \Box(\J_\nu \Rightarrow \Box \J_\nu)$ for all configurations $\C$ satisfying $\Gamma_\nu$. If we succeed, we put $\T_\nu = \T \cup \J_\nu$. Otherwise, $\T_\nu = \T \cup \Exp_\nu$. Intuitively, the set $\J_\nu$ is the largest subset $M$ of $\Exp_\nu$ such that every element of $M$ can be re-enabled only by firing a transition which has been identified as permanently disabled. This again leads to a simple fixed-point computation, which is detailed in the appendix.
\begin{toappendix}
\subsection{The procedure for computing $\J_\nu$.}
\label{app-J-set}	
Consider a subset $M \subseteq \Exp_\nu$ and a configuration $\C'$ reachable from a configuration satisfying $\Gamma_\nu$ and such that $\C' \models M$. Let $\C_0,\ldots,\C_\ell$ be an execution initiated in $\C'$ such that $\C_i \models M$ for all $i<\ell$, and some transition specified by $M$ is re-enabled in $\C_\ell$. Let $AB \tr CD$ be the transition fired when moving from $\C_{\ell-1}$ to $\C_{\ell}$. Since $\C_{\ell-1} \models M$ and firing $AB \tr CD$ enables some transition specified by $M$ in $\C_\ell$, one of the following conditions holds:
\begin{itemize}
  \item $CD \in M$,
  \item there is $E \in V$ such that $CE \in M$, $E \neq D$, $\C_{\ell-1}(E) > 0$,
  \item there is $E \in V$ such that $DE \in M$, $E \neq C$, $\C_{\ell-1}(E) > 0$.
\end{itemize}
The set $\J_\nu$ is the \emph{largest} $M \subseteq \Exp_\nu$ such that, for every $EF \in M$, the following holds:
\begin{itemize}
\item For every transition of the form $AB \tr EF$ the formula $(\pi_\nu \wedge \T \wedge M) 
	\Rightarrow \ \xi_{AB}$ is a propositional tautology.
\item For every transition of the form $AB \tr EG$ where $G \neq F$ we have that
   \begin{itemize}
   	  \item if $E \neq F$, then the formula $(\neg E \wedge F \wedge \pi_\nu \wedge \T \wedge M) \Rightarrow \ \xi_{AB}$ is a tautology;
   	  \item if $E = F$, then $A=E$, or $B = E$, or $(E! \wedge \pi_\nu \wedge \T \wedge M) \Rightarrow \ \xi_{AB}$ is a tautology.
   \end{itemize}
\item For every transition of the form $AB \tr FG$ where $G \neq E$ we have that
   \begin{itemize}
   	\item if $E \neq F$, then the formula $(\neg F \wedge E \wedge \pi_\nu \wedge \T \wedge M) \Rightarrow \ \xi_{AB}$ is a tautology;
	\item if $E = F$, then $A=F$, or $B = F$, or $(F! \wedge \pi_\nu \wedge \T \wedge M) \Rightarrow \ \xi_{AB}$ is a tautology.
   \end{itemize}
\end{itemize}
Observe that $\J_\nu$ is computable by a simple fixed-point algorithm. 
\end{toappendix}

A proof of the next lemma is straightforward. 

\begin{lemma}
	\label{lem-as-F}
	For every configuration $\C$ such that $\C \models \Gamma_\nu$ we have that 
	\begin{enumerate}
		\item[(a)] $\C \models \Diamond  \Box\big(\pi_\nu \wedge \T \wedge \Exp_\nu)$
		\item[(b)] $\C \models \Box(\J_\nu \ \Rightarrow \ \Box\J_\nu )$
	\end{enumerate}
\end{lemma}

\noindent
If $\J_\nu \neq \emptyset$, we
put $\T_\nu = \T \cup \J_\nu$. Otherwise, we put $\T_\nu = \T \cup \Exp_\nu$.
\bigskip

\noindent
\textbf{Computing $\Phi_\nu$.}\quad We say that a configuration $\C$ is \emph{$S_\nu$-entering} if $\C \models \Box\pi_\nu \wedge \Box \T_\nu$ and there is an execution $\C_0,\ldots,\C_\ell$ such that $\C_0 \models \Gamma_\nu$, $\C_\ell = \C$, and $\C_j \not\models \Box\pi_\nu \wedge \Box \T_\nu$ for all $j<\ell$. An immediate consequence of Lemma~\ref{lem-as-F} is the following:

\begin{lemma}
	Almost all runs initiated in a configuration satisfying $\Gamma_\nu$ visit an $S_\nu$-entering configuration.
\end{lemma}

\noindent
The formula $\Phi_\nu$ specifies the properties of \mbox{$S_\nu$-entering} configurations. The formula $\Phi_\nu$ always implies $\pi_\nu \wedge \T_\nu$, but it can also be more detailed if $\J_\nu \neq \emptyset$. More precisely, we say that $\J_\nu$ is \emph{$\nu$-disabled} if $\J_\nu \neq \emptyset$ and for all $AB \in \J_\nu$ we have that $\nu \Rightarrow \xi_{AB}$ is a propositional tautology (i.e., all transitions specified by $\J_\nu$ are disabled in all configurations satisfying~$\nu$). Similarly, $\J_\nu$ is \emph{$\nu$-enabled} if $\J_\nu \neq \emptyset$ and there exists $AB \in \J_\nu$ such that $\nu \Rightarrow \neg \xi_{AB}$ is a tautology (i.e., some transition specified by $\J_\nu$ is enabled in all configurations satisfying~$\nu$).

Observe that if $\J_\nu$ is $\nu$-disabled, then all transitions specified by $\J_\nu$ are simultaneously disabled in every configuration $\C$ satisfying $\Gamma_\nu$. Hence, all $S_\nu$-entering configurations satisfy $\Gamma_\nu$ (see Lemma~\ref{lem-as-F}~(b)). Now suppose that $\J_\nu$ is $\nu$-enabled, and let $Q_\nu$ be the set of all $A \in Q$ such that $AB \in \J_\nu$ for some $B \in Q$. Since for every configuration $\C$ satisfying $\Gamma_\nu$ there is a transition specified by $\J_\nu$ enabled in $\C$, the last transition executed before visiting an $S_\nu$-entering configuration must be a transition ``transforming'' some $A \in Q_\nu$, i.e., a transition of the form $AB \tr CD$ generating an edge $A \tran{} C$ of $G_\nu$. Let $\calK_\nu$ be the set of all right-hand sides of all such transitions.
The formula $\Phi_\nu$ is defined as follows:
\[
   \Phi_\nu \quad \equiv \quad
   \begin{cases}
       \displaystyle \pi_\nu \wedge \T_\nu \ \wedge \ \nu& 
                  \mbox{if $\J_\nu$ is $\nu$-disabled,}\\[1ex]
       \displaystyle\pi_\nu \wedge \T_\nu \ \wedge\ \bigg(\bigvee_{CD \in \calK_\nu} \neg\xi_{CD} \bigg) & \mbox{if $\J_\nu$ is $\nu$-enabled,}\\
       \pi_\nu \wedge \T_\nu& \mbox{otherwise.}
   \end{cases}
\]
It is easy to check that every $S_\nu$-entering configuration satisfies the formula~$\Phi_\nu$. The constructed $\calP$-stage $S_\nu = (\Phi_\nu,\pi_\nu,\T_\nu)$ becomes a successor of the $\calP$-stage~$S$.
 
\subsubsection{Computing $\T_\nu$ and $\Phi_\nu$: Case $\Exp_\nu = \emptyset$.}
\label{sec-Exp-empty}

In this case $G_\nu$ is a collection of disconnected SCCs. We put $\T_\nu = \T$. In the rest of the section we show how to construct the formula  $\Phi_\nu$. 

We say that an edge $A \tran{} B$ of $G_\nu$ is \emph{stable} if there is a transition $AC \tr BD$ generating $A \tran{} B$ such that $\pi_\nu(C) = \ttt$. Let $\I_\nu$ be the union of all non-bottom SCCs of the directed graph obtained from $G_\nu$ by considering only the stable edges of $G_\nu$.

\begin{lemma}
	\label{lem-disable}
	For every configuration $\C$ such that $\C \models \Gamma_\nu$ we have that 
	\mbox{$\C \models \Diamond \big( \bigwedge_{A \in \I_\nu} \neg A \big)$}.
\end{lemma}

\noindent
Similarly as above, we say that $\C$ is \emph{$S_\nu$-entering} if 
\(
 \C \ \models\  \Box\pi_\nu \ \wedge \ \Box \T_\nu \ \wedge \ \bigwedge_{A \in \I_\nu} \neg A
\)  
and there is an execution $\C_0,\ldots,\C_\ell$ such that $\C_0 \models \Gamma_\nu$, $\C_\ell = \C$, and $\C_j$ does \emph{not} satisfy the above formula for all $j<\ell$.

Observe that if $\nu(A) = \fff$ for all $A \in \I_\nu$, then $\nu$ implies $\bigwedge_{A \in \I_\nu} \neg A$ and hence every configuration $\C$ satisfying $\Gamma_\nu$ is $S_\nu$-entering. Further, if $\nu(A) = \ttt$ for some $A \in \I_\nu$, then the last transition executed before visiting an $S_\nu$-entering configuration is a transition $EF \tr CD$ generating a stable edge $E \tran{} C$ of $G_\nu$ where $E \in \I_\nu$ and $C \not\in \I_\nu$.  Let $\calL_\nu$ be the set of all right-hand sides of all such transitions. We put
\[
  \Phi_\nu \equiv \begin{cases}
                     \displaystyle\pi_\nu \wedge \T_\nu \wedge \ \bigg(\bigwedge_{A \in \I_\nu} \neg A\bigg) \wedge \ \bigg(\bigvee_{CD \in \calL_\nu} \neg\xi_{CD}\bigg)  &
                     \mbox{if $\nu(A) = \ttt$ for some $A \in \I_\nu$,}\\[2ex]
                      \displaystyle\pi_\nu \wedge \T_\nu \wedge \ \nu  &
 \mbox{if $\nu(A) = \fff$ for all $A \in \I_\nu$,}\\[2ex]
                     \displaystyle\pi_\nu \ \wedge \ \T_\nu \ \wedge \bigg(\bigwedge_{A \in \I_\nu} \neg A\bigg)   & \mbox{otherwise.}
                  \end{cases} 
\]

We say that the constructed $\calP$-stage $S_\nu = (\Phi_\nu,\pi_\nu,\T_\nu)$ is \emph{redundant} if there is a $\calP$-stage $S' = (\Phi',\pi',\T')$ on the path from the initial stage $S_0$ to $S$ such that $\pi_\nu = \pi'$, $\T_\nu = \T'$, and $\Phi'$ implies $\Phi_\nu$. The $\calP$-stage $S_\nu$ becomes a successor of $S$ iff $S_\nu$ is not redundant. This ensures termination of the algorithm even for poorly designed population protocols. 

\begin{toappendix}

\subsection{A pseudocode for computing the stage $S_\nu = (\Phi_\nu,\pi_\nu,\T_\nu)$}
\label{sec-pseudocode}
An explicit pseudocode for constructing the stage $S_\nu = (\Phi_\nu,\pi_\nu,\T_\nu)$ is given in Algorithm~\ref{alg-stage-compute}.
	
\begin{algorithm}
	\SetAlgoLined
	\DontPrintSemicolon
	\SetKwInOut{Input}{input}\SetKwInOut{Output}{output}
	\SetKwData{n}{n}\SetKwData{f}{f}\SetKwData{g}{g}
	\SetKwData{Low}{l}\SetKwData{x}{x}
	\Input{$S = (\Phi,\pi,\T)$, an assignment $\nu \in \Val_{S}$}
	\Output{$S_\nu = (\Phi_\nu,\pi_\nu,\T_\nu)$. }
	\BlankLine
	compute $\pi_\nu$ \; \label{l-DEsets}
	\If{$(\pi_\nu,\T)$ is stable or dead\label{l-if-start}}{\Return $(\pi_\nu,\pi_\nu,\T)$}\label{l-if-end}
	\BlankLine
	compute $G_\nu$ \; \label{l-Ggraph}
	compute $\Exp_\nu$ \; \label{l-Exp}
	compute $\J_\nu$ \; \label{l-Jcomp}
	\BlankLine
	\eIf{$\Exp_\nu \neq \emptyset$}{
		\eIf{$\J_\nu \neq \emptyset$}{\label{l-start}
			$\T_\nu := \T \cup \J_\nu$\;
			\uIf{$\J_\nu$ is $\nu$-disabled}{$\displaystyle\Phi_\nu := \pi_\nu \wedge \T_\nu \wedge \nu$ \;}
			\uElseIf{$\J_\nu$ is $\nu$-enabled}{
				compute the set $\calK_\nu$\;  
				$\displaystyle\Phi_\nu := \pi_\nu \wedge \T_\nu \ \wedge\ \bigg(\bigvee_{CD \in \calK_\nu} \eta(CD) \bigg)$ \;}
			\Else{$\Phi_\nu := \pi_\nu \wedge \T_\nu$}
		}{$\T_\nu := \T \cup \Exp_\nu$\; $\Phi_\nu := \pi_\nu \wedge \T_\nu$ \;}\label{l-end}
	}{
		$\T_\nu := \T$\;
		compute the set $\I_\nu$ \; \label{l-Iset}
		\uIf{$\nu(A) = \ttt$ for some $A \in \I_\nu$}{
		   compute the set $\calL_\nu$ \; 
		   $\displaystyle\Phi_\nu := \pi_\nu \wedge \T_\nu\wedge  \bigg(\bigwedge_{A \in \I_\nu} \neg A\bigg) \wedge \bigg(\bigvee_{CD \in \calL_\nu} \eta(CD)\bigg)$ \;
	    }
        \uElseIf{$\nu(A) = \fff$ for all $A \in \I_\nu$}{
           $\displaystyle\Phi_\nu := \pi_\nu \wedge \T_\nu\wedge \nu$ \;
        }
        \Else{$\displaystyle\Phi_\nu := \pi_\nu \wedge \T_\nu\wedge  \bigg(\bigwedge_{A \in \I_\nu} \neg A\bigg)$}
	}
	\BlankLine	
	\Return $(\Phi_\nu,\pi_\nu,\T_\nu)$ \;
	\caption{Computing the $\calP$-stage $S_\nu$.}
	\label{alg-stage-compute}
\end{algorithm}
\end{toappendix}

%% Time
\section{Computing the interaction complexity}
\label{sec-expected}

We show how to compute an upper asymptotic bounds on $\Next_S$ for every stage $S$ in the stage graph constructed in Section~\ref{sec-stage-compute}.

For the rest of this section, we fix a population protocol $\calP = (Q,T,\Sigma,I,O)$, a $\calP$-stage $S = (\Phi,\pi,\T)$, and its successor $S_\nu = (\Phi_\nu,\pi_\nu,\T_\nu)$. Recall the formula $\Gamma_\nu$, the graph $G_\nu = (V,{\tran{}})$, and the sets $\Exp_\nu$, $\J_\nu$ defined in Section~\ref{sec-stage-compute}. We show how to compute an asymptotic upper bound on the function $\Reach_{S,S_\nu}$ that assigns to every $n \geq 1$ the maximal $\Ex_{\C}[\Time_{\Enter(S_\nu)}]$, where $\Enter(S_\nu)$ is a fresh atomic proposition satisfied precisely by all $S_\nu$-entering configurations, and $\C$ ranges over all configurations of size~$n$ satisfying $\Gamma_\nu$ (if there is no such configuration of size $n$, we put $\Reach_{S,S_\nu}(n) = 0$).  Observe that $\max_{S_\nu}\{\Reach_{S,S_\nu}\}$, where $S_\nu$ ranges over all successor stages of $S$, is then an asymptotic upper bound on $\Next_S$.

Let us note that if $\calP$ terminates, then $\Comp_\calP \in 2^{2^{\calO(n)}}$. This trivial bound follows by observing that the number of all configurations of size $n$ is $2^{\calO(n)}$, and the probability of reaching a stable configuration in  $2^{\calO(n)}$ transitions is $2^{-2^{o(n)}}$; this immediately implies the mentioned upper bound on $\Comp_\calP$.  As we shall see, the worst asymptotic bound on $\Reach_{S,S_\nu}$ is $2^{\calO(n)}$, and in many cases, our results allow to derive even a polynomial upper bound on $\Reach_{S,S_\nu}$.

Recall that if $(\pi_\nu,\T)$ is stable or dead, we have that $\Reach_{S,S_\nu}(n) = 0$ for all $n \in \Nset$ (in this case, we define $S_\nu$-entering configurations are the configurations satisfying $\Box(\pi_\nu \wedge \T)$).  Now suppose  $(\pi_\nu,\T)$ is not stable or dead. Furthermore, let us first assume $\Exp_\nu = \emptyset$. Then, the upper bound on $\Reach_{S,S_\nu}$ is singly exponential in~$n$. 
\begin{theoremrep}
	If $\Exp_\nu = \emptyset$, then $\Reach_{S,S_\nu} \in 2^{\calO(n)}$.
\end{theoremrep}

\begin{proof}
	Recall the definition of $\I_\nu$ given in Section~\ref{sec-Exp-empty}.	Let $\C$ be a configuration of size~$n$ reachable from a configuration satisfying the formula $\Gamma_\nu$. Then there is a configuration $\C'$ reachable from $\C$ in at most $|Q|\cdot n$ transitions such that $\C' \models  \bigwedge_{A \in \I_\nu} \neg A$. The probability of firing a given transition in a given configuration is at least $1/n^2$, hence the probability of reaching such a $\C'$ from $\C$ in at most $|Q|\cdot n$ transitions is $2^{-\calO(n)}$. On average, we need to perform such an execution at most $2^{\calO(n)}$ times, which yields the $2^{\calO(n)}$ bound.    
\end{proof}	

Now assume $\Exp_\nu \neq \emptyset$.  Let $\U \subseteq Q$ be the set of all states appearing in some non-bottom SCC of~$G_\nu$. We start with some auxiliary definitions.

\begin{definition}
For every $A \in \U$, let $\Exp_\nu[A]$ be the set of all $B \in Q$ such that $AB \in \Exp_\nu$. We say that $S_\nu$ is \emph{fast} if, for every $A \in \U$, the formula
\(
   \big( \pi_\nu \wedge \T \wedge \neg \Exp_{\nu} \wedge A \big)\ \Rightarrow \ \big( \bigvee_{B \in \Exp_\nu[A]} \neg\xi_{AB} \big)
\)
is a propositional tautology.
\end{definition}

\begin{definition}
\label{def-very-fast}
For every $A \in V$, let $[A]$ be the SCC of $G_\nu$ containing $A$. We say that $S_\nu$ is \emph{very fast} if every transition $AB \tr CD$ such that $AB,CD \in V^{\ms{2}}$ and $\{A,B,C,D\} \cap \U \neq \emptyset$ satisfies one of the following conditions:
\begin{itemize}
	\item The formula $\big( \pi_\nu \wedge \T \big)\ \Rightarrow \ \xi_{AB}$ is a propositional tautology.
	\item $[C] \neq [A] \neq [D]$ and $[C] \neq [B] \neq [D]$.
\end{itemize}  
\end{definition}

\begin{theoremrep}
\label{thm-J-non-empty}
If $\Exp_\nu \neq \emptyset$ and $\J_\nu \neq \emptyset$, then 
\begin{itemize}
\item $\Reach_{S,S_\nu} \in \calO(n^3)$. 
\item If  $S_\nu$ is fast, then $\Reach_{S,S_\nu} \in \calO(n^2 \cdot \log(n))$.
\item If  $S_\nu$ is very fast, then $\Reach_{S,S_\nu} \in \calO(n^2)$.
\end{itemize}
\end{theoremrep}
\begin{proof}
	
For every SCC of $G_\nu$, we define its \emph{distance} inductively as follows: the distance of every bottom SCC is $0$, and the distance of a non-bottom SCC is the maximal distance of its immediate successors plus~$1$. For all $A \in Q$, let $w(A)$ be a non-negative integer defined by
\[
w(A) = \begin{cases}
d & \mbox{$A$ appears in a non-bottom SCC of $G$ with distance $d$,}\\
0 & \mbox{otherwise.}
\end{cases}
\] 
Since $\J_\nu \subseteq \Exp_\nu$, for every $n \in \Nset$ we have that $\Reach_{S,S_\nu}(n) \leq \max_\C \Ex_{\C}[\Time_{\Exp_{\nu}}]$, where $\C$ ranges over all configurations of size $n$ satisfying the formula $\Gamma_\nu$. Hence, it suffices to give an appropriate upper bound on $\Ex_{\C}[\Time_{\Exp_{\nu}}]$.
 	
Let $\C$ be a configuration of size $n$ reachable from a configuration satisfying the formula $\Gamma_\nu$. The \emph{potential} of $\C$ is defined by \mbox{$\alpha_{\C} = \sum_{A \in Q} w(A) \cdot \C(A)$}. Clearly, $0 \leq \alpha_{\C} \leq |Q| \cdot n$. Suppose that $\C$ fires a transition $AC \tr BD$ and enters a configuration $\C'$. It follows immediately from the definition of $G_\nu$ that $\alpha_{\C'} \leq \alpha_{\C}$. Further, $\alpha_{\C'} < \alpha_{\C}$ iff $AC \tr BD$ generates an edge $A \tran{} B$ of $G$ such that $A$ and $B$ belong to different SCC's of~$G_\nu$. Consequently, if $\alpha_\C = 0$,  then $\C \models \Exp_\nu$. We show that if $\C \not\models \Exp_\nu$, then the expected number of transitions fired before reaching a configuration $\C'$ such that $\C' \models \Exp_\nu$ or $\alpha_{\C'} < \alpha_{\C}$ is bounded by $c \cdot n^2$, where $c$ is a positive constant depending only of $\calP$. If $S_\nu$ is fast, then this bound can be improved to $c' \cdot (n^2 / \alpha_\C)$. Since $\alpha_\C \leq |Q| \cdot n$, we immediately obtain that $\Ex_{\C}[\Time_{\Exp_{\nu}}]$ is $\calO(n^3)$. If $S_\nu$ is fast, this bound can be improved to $\sum_{k=1}^{|Q|\cdot n} c' \cdot (n^2 / k) = c' \cdot n^2 \cdot \calH_{|Q|\cdot n}$, where $\calH_i$ is the $i$-th Harmonic number. Since $\calH_i$ is $\Theta(\log i)$, we obtain that $\Ex_{\C}[\Time_{\Exp_{\nu}}]$ is $\calO(n^2 \cdot \log(n))$.

So, let $\C$ be a configuration reachable from a configuration satisfying the formula $\Gamma_\nu$ such that $\C \not\models \Exp_\nu$. The probability of firing a transition leading to a $\C'$ such that either $\alpha_{\C'} < \alpha_{\C}$ or $\C' \models \Exp_\nu$ is at least $c/n^2$, where $c$ is a constant depending only on $\calP$. If $S_\nu$ is fast, then this bound can be improved to $c' \cdot (\alpha_\C/n^2)$ where $c'$ is another constant depending only on $\calP$ (this follows by observing that there is $A \in \U$ such that $\C(A) \geq  \alpha_{\C}/|Q|^2$).  If this trial is \emph{unsuccessful}, i.e., $\C$ executes a transition leading to a $\C'$ such that $\alpha_{\C'} = \alpha_\C$ and $\C' \not\models \Exp_\nu$, another \emph{independent} trial is performed in $\C'$ (the success probability is again at least $c/n^2$, or at least $c' \cdot (\alpha_{\C} /n^2)$ if $S_\nu$ is fast). Hence, on average, at most $n^2/c$ trials are needed to enter a configuration $\C''$ such that $\alpha_{\C''} < \alpha_{\C}$ or $\C'' \models \F_\nu$. If $S_\nu$ is fast, this bound can be improved to $c'' \cdot (n^2/\alpha_{\C})$, where $c'' = 1/c'$.

The case when $S_\nu$ is very fast is handled similarly. For every configuration $\C$ reachable form a configuration satisfying $\Gamma_\nu$, we say that a given $A \in V$ is \emph{active} in $\C$ if there is a transition of the form $AB \tr CD$ enabled in $\C$. Let $\ta_{\C}$ be the set of all active $A$'s for which there is no active $B$  such that $[A] \neq [B]$ and $[A]$ is reachable from $[B]$ in the graph of SCC's determined by $G_\nu$. Let $\beta_\C = \sum_{A \in \ta_{\C}} d([A])$, where $d([A])$ is the distance of $[A]$. Note that if $\beta_\C = 0$, then $\C \models \Exp_\nu$. We show that if $\beta_\C > 0$, then the expected number of transition fired before reaching a configuration $\C'$ such that  $\beta_{\C'} < \beta_{\C}$ is $\calO(n^2)$. This clearly suffices, because $\beta_\C \leq |Q|^2$. First, observe that there must be $A,B \in \ta_{\C}$ and a transition of the form $AB \tr CD$ enabled in $\C$. Since $A,B \in \ta_{\C}$, the number of $A$'s and $B$' can only decrease along all executions initiated in $\C$ (see Definition~\ref{def-very-fast}), and the transition $AB \tr CD$ can be fired at most $\min\{\C(A),\C(B)\}$ times. If $\C'$ is a successor of $\C$ such that $\min\{\C'(A),\C'(B)\} =i$, the probability of firing  $AB \tr CD$ in $\C'$ is at least $i^2/(c \cdot n^2)$ where $c$ is the total number of transitions with the head~$AB$. Since $\min\{\C(A),\C(B)\} \leq n$, the expected number of transition needed to enter a configuration $\C''$ from $\C$ such that $\C''(A) = 0$ or $\C''(B) = 0$ is bounded by $\sum_{i=1}^n (c \cdot n^2)/i^2 = c \cdot n^2 \cdot \calH_{n,2} \in \calO(n^2)$. It is easy to check that $\beta_{\C''} < \beta_{\C}$, and we are done. 
\end{proof}	

Computing an asymptotic upper bound on $\Reach_{S,S_\nu}$ when $\Exp_\nu \neq \emptyset$ and $\J_\nu = \emptyset$ is more complicated. We show that a \emph{polynomial} upper bound always exists, and that the degree of the polynomial is computable. However, our proof does not yield an efficient algorithm for computing/estimating the degree.
   
\begin{theoremrep}
	If $\Exp_\nu \neq \emptyset$ and $\J_\nu = \emptyset$, then $\Reach_{S,S_\nu} \in \calO(n^c)$ for some computable constant~$c$.
\end{theoremrep}
\begin{proof}
	Let $\C$ be a configuration of size $n$ reachable from a configuration satisfying the formula $\Gamma_\nu$. By using the arguments of Theorem~\ref{thm-J-non-empty}, we obtain that $\Ex_{\C}[\Time_{\Exp_{\nu}}]$ is $\calO(n^3)$. However, after reaching a configuration $\C'$ such that $\C' \models \Exp_\nu$, it may happen that the transitions specified by $\Exp_\nu$ are disabled only temporarily, i.e., it is still possible to reach a configuration $\C''$ from $\C'$ such that $\C'' \not\models \Exp_\nu$. First, we show that if such a $\C''$ is reachable from $\C'$, then it is reachable in at most $d$ transitions, where $d$ is a constant depending only on $\calP$. This follows by observing that the set of configurations which can reach such a $\C''$ is upward closed w.r.t.{} point-wise ordering, and hence there (effectively) exist finitely many \emph{minimal} configurations with this property. Each of these minimal configurations can reach a configuration violating $\Exp_\nu$ in a constant number of transitions, and all larger configurations can perform the same sequence and thus reach a configuration violating $\Exp_\nu$. Hence, the $d$ can be chosen as the maximum of these finitely many (computable) constants. 
	
    A \emph{progress transition} is a transition of the form $AB \tr CD$ where $AB \in \Exp_{\nu}$ and $AB \tr CD$ generates an edge of $G_\nu$ connecting two different SCCs of $G_\nu$. Note that the total number of progress transitions fired along a run initiated in $\C'$ is bounded by $|Q|\cdot n$. Furthermore, the probability of executing a progress transition in at most $d+1$ steps is bounded from below by  $n^{-2(d+1)}$. Hence, on average, we need to perform at most $n^{2(d+1)}$ executions of length $d+1$ to fire a progress transition or reach a configuration satisfying $\Box \Exp_\nu$. This implies  $\Reach_{S,S_\nu}(n)$ is bounded by $|Q|\cdot n \cdot n^{2(d+1)} \cdot (d+1)$, which is $\calO(n^c)$ where $c = 2d + 3$. 
\end{proof}

%% Experimental
\section{Experimental results}
\label{sec-experiments}

We have implemented our approach as a tool\footnote{The tool and its
  benchmarks are available at
  \url{https://github.com/blondimi/pp-time-analysis}.} that takes a
population protocol as input and follows the procedure of
Section~\ref{sec-stage-compute} to construct a stage graph together with an
upper bound on $\Comp_\calP$. Our tool is implemented in
\textsc{Python 3}, and uses the SMT solver \textsc{Microsoft
  Z3}~\cite{MB08} to test for tautologies and to obtain valid
valuations.

\begin{table}[!h]
  \centering
  \begin{minipage}[t]{0.46225\linewidth}
    \vspace{0pt}\resizebox{\linewidth}{!}{
      \begin{tabular}{|l|r|r|r|c|r|}
      \hline

      \multicolumn{3}{|c|}{\rule{0pt}{10pt}\textbf{Protocol}} &
      \multicolumn{1}{c|}{\multirow{2}{*}{\rule{0pt}{10pt}$|\calS|$}} &
      \multicolumn{1}{c|}{\multirow{2}{*}{\rule{0pt}{10pt}\textbf{Bound}}} &
      \multicolumn{1}{c|}{\multirow{2}{*}{\rule{0pt}{10pt}\textbf{Time}}} \\

      \cline{1-3}
      
      \multicolumn{1}{|c|}{predicate / params.} &
      \multicolumn{1}{c|}{$|Q|$} & \multicolumn{1}{c|}{$|T|$} &
      \multicolumn{1}{c|}{} & \multicolumn{1}{c|}{} &
      \multicolumn{1}{c|}{} \\

      \hline

      $x_1 \lor \ldots \lor x_n$~\cite{guidelines} & 2 & 1 & 5 & $n^2
      \cdot \log n$ & 0.1 \\

      $x \geq y$~(Example~\ref{exa-pop-ninja4}) & 5 & 6 & 13 & $n^2 \cdot \log
      n$ & 0.4 \\

      $x \geq y$~\tablefootnote{Protocol of Example~\ref{exa-pop-long}
        without the tie-breaking rule $ba \mapsto bb$ (only correct if
        $x \neq y$).} & 4 & 3 & 9 & $n^2 \cdot \log n$ & 0.2 \\

      $x \geq y$~(Example~\ref{exa-pop-long}) & 4 & 4 & 11 & $\mathrm{exp}(n)$
      & 0.3 \\      

      \hline \multicolumn{6}{|l|}{\rule{0pt}{10pt}Flocks-of-bird
        protocol~\cite{DBLP:journals/dc/AngluinADFP06}: $x \geq c$} \\ \hline

      $c = 5$ & 6 & 21 & 26 & $n^3$ & 0.8 \\
      $c = 10$ & 11 & 66 & 46 & $n^3$ & 4.0 \\
      $c = 15$ & 16 & 136 & 66 & $n^3$ & 12.1 \\
      $c = 20$ & 21 & 231 & 86 & $n^3$ & 28.9 \\
      $c = 25$ & 26 & 351 & 106 & $n^3$ & 58.0 \\
      $c = 30$ & 31 & 496 & 126 & $n^3$ & 118.9 \\
      $c = 35$ & 36 & 666 & 146 & $n^3$ & 222.3 \\
      $c = 40$ & 41 & 861 & 166 & $n^3$ & 366.2 \\
      $c = 45$ & 46 & 1081 & 186 & $n^3$ & 495.3 \\
      $c = 50$ & 51 & 1326 & 206 & $n^3$ & 952.8 \\
      $c = 55$ & 56 & 1596 & --- & --- & \textsc{T/O} \\

      \hline

      \multicolumn{6}{|l|}{\rule{0pt}{10pt}Logarithmic flock-of-birds
        protocol\tablefootnote{An adapted version of the protocol
          of~\cite[Sect.~3]{BEJ18} without so-called $k$-way
          transitions.}: $x \geq c$} \\ \hline

      $c = 15$ & 8 & 23 & 66 & $n^3$ & 2.6 \\
      $c = 31$ & 10 & 34 & 130 & $n^3$ & 6.1 \\
      $c = 63$ & 12 & 47 & 258 & $n^3$ & 13.9 \\
      $c = 127$ & 14 & 62 & 514 & $n^3$ & 39.4 \\
      $c = 255$ & 16 & 79 & 1026 & $n^3$ & 81.0 \\
      $c = 1023$ & 20 & 119 & 4098 & $n^3$ & 395.7 \\
      $c = 2047$ & 22 & 142 & 8194 & $n^3$ & 851.9 \\
      $c = 4095$ & 24 & 167 & --- & --- & \textsc{T/O} \\

      \hline

      \end{tabular}}
    \end{minipage}\hspace{1pt}
    \begin{minipage}[t]{0.5\linewidth}
    \vspace{0pt}\resizebox{\linewidth}{!}{
    \begin{tabular}{|l|r|r|r|c|r|}
      \hline

      \multicolumn{3}{|c|}{\rule{0pt}{10pt}\textbf{Protocol}} &
      \multicolumn{1}{c|}{\multirow{2}{*}{\rule{0pt}{10pt}$|\calS|$}} &
      \multicolumn{1}{c|}{\multirow{2}{*}{\rule{0pt}{10pt}\textbf{Bound}}} &
      \multicolumn{1}{c|}{\multirow{2}{*}{\rule{0pt}{10pt}\textbf{Time}}} \\

      \cline{1-3}
      
      \multicolumn{1}{|c|}{predicate / params.} &
      \multicolumn{1}{c|}{$|Q|$} & \multicolumn{1}{c|}{$|T|$} &
      \multicolumn{1}{c|}{} & \multicolumn{1}{c|}{} &
      \multicolumn{1}{c|}{} \\

      \hline
      
      \multicolumn{6}{|l|}{\rule{0pt}{10pt}Flocks-of-bird
        protocol~\cite{guidelines}: $x \geq c$} \\ \hline

      $c = 5$ & 6 & 9 & 54 & $n^3$ & 2.5 \\
      $c = 7$ & 8 & 13 & 198 & $n^3$ & 11.3 \\
      $c = 10$ & 11 & 19 & 1542 & $n^3$ & 83.9 \\
      $c = 13$ & 14 & 25 & 12294 & $n^3$ & 816.4 \\
      $c = 15$ & 16 & 29 & --- & --- & \textsc{T/O} \\

      \hline
      
      \multicolumn{6}{|l|}{\rule{0pt}{10pt}Average-and-conquer
        protocol\tablefootnote{The protocol is only correct assuming
          $x \neq y$.}~\cite{AGV15}: $x \geq y$ with params.\ $m$ and
        $d$} \\ \hline
      
      $m = 3, d = 1$ & 6 & 21 & 41 & $n^2 \cdot \log n$ & 2.0 \\
      $m = 3, d = 2$ & 8 & 36 & 1948 & $n^2 \cdot \log n$ & 98.7 \\
      $m = 5, d = 1$ & 8 & 36 & 1870 & $n^3$ & 80.1 \\
      $m = 5, d = 2$ & 10 & 55 & --- & --- & \textsc{T/O} \\
      $m = 7, d = 1$ & 10 & 55 & --- & --- & \textsc{T/O} \\

      \hline
      
      \multicolumn{6}{|l|}{\rule{0pt}{10pt}Remainder
        protocol~\cite{DBLP:conf/podc/BlondinEJM17}: $\sum_{1 \leq i <
          m} i \cdot x_i \equiv 0\ (\mathrm{mod}\ m)$} \\ \hline

      $m = 3$ & 5 & 12 & 27 & $n^2 \cdot \log n$ & 0.8 \\
      $m = 5$ & 7 & 25 & 225 & $n^2 \cdot \log n$ & 12.5 \\
      $m = 7$ & 9 & 42 & 1351 & $n^2 \cdot \log n$ & 88.9 \\
      $m = 9$ & 11 & 63 & 7035 & $n^2 \cdot \log n$ & 544.0 \\
      $m = 10$ & 12 & 75 & --- & --- & \textsc{T/O} \\

      \hline \multicolumn{6}{|l|}{\rule{0pt}{10pt}Threshold
        protocol~\cite{DBLP:journals/dc/AngluinADFP06}: $\sum_{1 \leq
          i \leq k} a_i \cdot x_i < c$} \\ \hline

      $-x_1 + x_2 < 0$ & 12 & 57 & 21 & $n^3$ & 3.0 \\
      $-x_1 + x_2 < 1$ & 20 & 155 & 131 & $n^3$ & 30.3 \\
      $-x_1 + x_2 < 2$ & 28 & 301 & --- & --- & \textsc{T/O} \\
      $-2x_1 - x_2 + x_3 + 2x_4 < 0$ & 20 & 155 & 1049 & $n^3$ & 166.3 \\
      $-2x_1 - x_2 + x_3 + 2x_4 < 1$ & 20 & 155 & 1049 & $n^3$ & 155.2 \\
      $-2x_1 - x_2 + x_3 + 2x_4 < 2$ & 28 & 301 & --- & --- & \textsc{T/O} \\
      
      \hline
    \end{tabular}}
    \end{minipage}
    \vspace{5pt}
    
    \caption{Results of the experimental evaluation where $|Q|$, $|T|$
      and $|\calS|$ correspond respectively to the number of states
      and transitions of the protocol, and the number of nodes of its
      stage graph.} \label{fig:benchmarks}
\end{table}

We tested our implementation on multiple protocols drawn from the
literature: a simple broadcast protocol~\cite{guidelines}, the
majority protocols of Example~\ref{exa-pop-long},
Example~\ref{exa-pop-ninja4} and~\cite{AGV15}, various flock-of-birds
protocols~\cite{DBLP:journals/dc/AngluinADFP06,guidelines,BEJ18}, a
remainder protocol~\cite{DBLP:conf/podc/BlondinEJM17} and a threshold
protocol~\cite{DBLP:journals/dc/AngluinADFP06}. Most of these
protocols are parametric, i.e.\ they are a family of protocols
depending on some parameters. For these protocols, we increased their
parameters until reaching a timeout. In particular, for the
logarithmic flock-of-birds protocol computing $x \geq c$, we used
thresholds of the form $c = 2^i - 1$ as they essentially consist the
most complicated case of the protocol.

All tests were performed on the same computer equipped
with eight Intel® Core™ i5-8250U 1.60 GHz CPUs, 8 GB of memory and
Ubuntu Linux 17.10 (64 bits). Each test had a timeout of 1000 seconds
(${\sim}16.67$ minutes). The duration of each test was evaluated as
the sum of the \texttt{user} time and \texttt{sys} time reported by
the \textsc{Python} time library.

The results of the benchmarks are depicted in
Table~\ref{fig:benchmarks}, where the \emph{bound} column refers to
the derived upper bound on $\Comp_\calP$. In particular, the tool
derived exponential and $n^2 \cdot \log n$ bounds for the protocols of
Example~\ref{exa-pop-long} and Example~\ref{exa-pop-ninja4}
respectively. The generated trees across all instances grow in width
but not much in height: the maximum height between the roots and the
leaves varies between $2$ and $5$, and most nodes are leaves.

It is worth noting that the $n^2 \log n$ bounds obtained in
Table~\ref{fig:benchmarks} for the \emph{average-and-conquer} and
\emph{remainder} protocols are tight with respect to the best known
bounds~\cite{AGV15,DBLP:journals/dc/AngluinADFP06}. However, some of
the obtained bounds are not tight, e.g.~we report $n^3$ for the
\emph{threshold protocol} but an $n^2 \log n$ upper bound was shown
in~\cite{DBLP:journals/dc/AngluinADFP06}. Moreover, it seems possible
to decrease the $n^3$ bound to $n^2$ for the \emph{flocks-of-bird
  protocol} of~\cite{DBLP:journals/dc/AngluinADFP06}. We are unsure of
the precise bounds for the remaining protocols.

\begin{toappendix}
\subsection*{Detailed experimental results}

\begin{longtabu}[c]{|l|r|r|r|r|c|c|r|}
      \hline
      \multicolumn{3}{|c|}{\rule{0pt}{10pt}\textbf{Protocol}} &
      \multicolumn{3}{c|}{\rule{0pt}{10pt}\textbf{Stage tree}} &
      \multicolumn{2}{c|}{\rule{0pt}{10pt}\textbf{Results}} \\ \hline
      
      \multicolumn{1}{|c|}{predicate and parameters} &
      \multicolumn{1}{c|}{$|Q|$} & \multicolumn{1}{c|}{$|T|$} &
      \multicolumn{1}{c|}{\# stages} & \multicolumn{1}{c|}{\#
        leaves} & depth & bound & \multicolumn{1}{c|}{time (secs.)}
      \\ \hline

      $x_1 \lor \ldots \lor x_n$~\cite{guidelines} & 2 & 1 & 5 & 3 & 2
      & $n^2 \cdot \log n$ & 0.103 \\

      $x \geq y$~\footnote{Protocol of Example~\ref{exa-pop-ninja4}.}
      & 5 & 6 & 13 & 8 & 3 & $n^2 \cdot \log n$ & 0.375 \\

      $x \geq y$~\footnote{Protocol of Example~\ref{exa-pop-long}
        without the tie-breaking rule $ba \mapsto bb$ (only correct if
        $x \neq y$).} & 4 & 3 & 9 & 5 & 3 & $n^2 \cdot \log n$ & 0.221
      \\

      $x \geq y$~\footnote{Protocol of Example~\ref{exa-pop-long}.} &
      4 & 4 & 11 & 6 & 3 & $\mathrm{exp}(n)$ & 0.263 \\

      \hline \multicolumn{8}{|l|}{\rule{0pt}{10pt}Flocks-of-bird
        protocol~\cite{DBLP:journals/dc/AngluinADFP06}: $x \geq c$} \\ \hline

      $c = 2$ & 3 & 6 & 12 & 9 & 2 & $n^3$ & 0.268 \\      
      $c = 3$ & 4 & 10 & 18 & 15 & 2 & $n^3$ & 0.423 \\
      $c = 4$ & 5 & 15 & 22 & 19 & 2 & $n^3$ & 0.597 \\
      $c = 5$ & 6 & 21 & 26 & 23 & 2 & $n^3$ & 0.798 \\
      $c = 10$ & 11 & 66 & 46 & 43 & 2 & $n^3$ & 3.974 \\
      $c = 15$ & 16 & 136 & 66 & 63 & 2 & $n^3$ & 12.121 \\
      $c = 20$ & 21 & 231 & 86 & 83 & 2 & $n^3$ & 28.945 \\
      $c = 25$ & 26 & 351 & 106 & 103 & 2 & $n^3$ & 58.022 \\
      $c = 30$ & 31 & 496 & 126 & 123 & 2 & $n^3$ & 118.855 \\
      $c = 35$ & 36 & 666 & 146 & 143 & 2 & $n^3$ & 222.251 \\
      $c = 40$ & 41 & 861 & 166 & 163 & 2 & $n^3$ & 366.247 \\
      $c = 45$ & 46 & 1081 & 186 & 183 & 2 & $n^3$ & 495.266 \\
      $c = 50$ & 51 & 1326 & 206 & 203 & 2 & $n^3$ & 952.841 \\
      $c = 55$ & 56 & 1596 & --- & --- & --- & --- & \textsc{timeout} \\

      \hline \multicolumn{8}{|l|}{\rule{0pt}{10pt}Flocks-of-bird
        protocol~\cite{guidelines}: $x \geq c$} \\ \hline

      $c = 2$ & 3 & 3 & 12 & 9 & 2 & $n^3$ & 0.256 \\
      $c = 3$ & 4 & 5 & 18 & 15 & 2 & $n^3$ & 0.424 \\
      $c = 4$ & 5 & 7 & 30 & 27 & 2 & $n^3$ & 0.746 \\
      $c = 5$ & 6 & 9 & 54 & 51 & 2 & $n^3$ & 2.541 \\
      $c = 7$ & 8 & 13 & 198 & 195 & 2 & $n^3$ & 11.343 \\
      $c = 10$ & 11 & 19 & 1542 & 1539 & 2 & $n^3$ & 83.862 \\
      $c = 13$ & 14 & 25 & 12294 & 12291 & 2 & $n^3$ & 816.432 \\
      $c = 15$ & 16 & 29 & --- & --- & --- & --- & \textsc{timeout} \\
      
      \hline \multicolumn{8}{|l|}{\rule{0pt}{10pt}Remainder
        protocol~\cite{DBLP:conf/podc/BlondinEJM17}: $\sum_{1 \leq i <
          m} i \cdot x_i \equiv 0\ (\mathrm{mod}\ m)$} \\ \hline

      $m = 2$ & 4 & 7 & 7 & 3 & 3 & $n^2 \cdot \log n$ & 0.198 \\
      $m = 3$ & 5 & 12 & 27 & 14 & 3 & $n^2 \cdot \log n$ & 0.811 \\
      $m = 4$ & 6 & 18 & 79 & 45 & 3 & $n^2 \cdot \log n$ & 4.062 \\
      $m = 5$ & 7 & 25 & 225 & 134 & 3 & $n^2 \cdot \log n$ & 12.479 \\
      $m = 7$ & 9 & 42 & 1351 & 846 & 3 & $n^2 \cdot \log n$ & 88.856 \\
      $m = 9$ & 11 & 63 & 7035 & 4502 & 3 & $n^2 \cdot \log n$ & 543.931 \\
      $m = 10$ & 12 & 75 & --- & --- & --- & --- & \textsc{timeout} \\

      \hline

      \multicolumn{8}{|l|}{\rule{0pt}{10pt}Average-and-conquer
        protocol~\cite{AGV15}: $x \geq y$ with parameters $m$ and $d$,
        assuming $x \neq y$} \\ \hline

      $m = 3, d = 1$ & 6 & 21 & 41 & 25 & 3 & $n^2 \cdot \log n$ & 1.982 \\
      $m = 3, d = 2$ & 8 & 36 & 1948 & 1038 & 5 & $n^2 \cdot \log n$ & 98.711 \\
      $m = 5, d = 1$ & 8 & 36 & 1870 & 1119 & 4 & $n^3$ & 80.097 \\
      $m = 5, d = 2$ & 10 & 55 & --- & --- & --- & --- & \textsc{timeout} \\
      $m = 7, d = 1$ & 10 & 55 & --- & --- & --- & --- & \textsc{timeout} \\
      $m = 3, d = 3$ & 10 & 55 & --- & --- & --- & --- & \textsc{timeout} \\

      \hline \multicolumn{8}{|l|}{\rule{0pt}{10pt}Threshold
        protocol~\cite{DBLP:journals/dc/AngluinADFP06}: $\sum_{1 \leq
          i \leq k} a_i \cdot x_i < c$} \\ \hline

      $-x_1 + x_2 < 0$ & 12 & 57 & 21 & 14 & 3 & $n^3$ & 3.012 \\
      $-x_1 + x_2 < 1$ & 20 & 155 & 131 & 104 & 3 & $n^3$ & 30.314 \\
      $-x_1 + x_2 < 2$ & 28 & 301 & --- & --- & --- & --- & \textsc{timeout} \\
      $-2x_1 - x_2 + x_3 + 2x_4 < 0$ & 20 & 155 & 1049 & 834 & 3 &
      $n^3$ & 166.283 \\
      $-2x_1 - x_2 + x_3 + 2x_4 < 1$ & 20 & 155 & 1049 & 834 & 3 &
      $n^3$ & 155.238 \\
      $-2x_1 - x_2 + x_3 + 2x_4 < 2$ & 28 & 301 & --- & --- & --- &
      --- & \textsc{timeout} \\     

      \hline \multicolumn{8}{|l|}{\rule{0pt}{10pt}Logarithmic
        flock-of-birds protocol\footnote{An adapted
          version of the protocol of~\cite[Sect.~3]{BEJ18} without
          so-called $k$-way transitions.}: $x \geq c$} \\ \hline

      $c = 3$ & 4 & 7 & 18 & 15 & 2 & $n^3$ & 0.571 \\
      $c = 7$ & 6 & 14 & 34 & 31 & 2 & $n^3$ & 1.926 \\
      $c = 15$ & 8 & 23 & 66 & 63 & 2 & $n^3$ & 2.605 \\
      $c = 31$ & 10 & 34 & 130 & 127 & 2 & $n^3$ & 6.144 \\
      $c = 63$ & 12 & 47 & 258 & 255 & 2 & $n^3$ & 13.909 \\
      $c = 127$ & 14 & 62 & 514 & 511 & 2 & $n^3$ & 39.382 \\
      $c = 255$ & 16 & 79 & 1026 & 1023 & 2 & $n^3$ & 81.000 \\
      $c = 1023$ & 20 & 119 & 4098 & 4095 & 2 & $n^3$ & 395.650 \\
      $c = 2047$ & 22 & 142 & 8194 & 8191 & 2 & $n^3$ & 851.861 \\
      $c = 4095$ & 24 & 167 & --- & --- & --- & --- & \textsc{timeout} \\      
      
      \hline
  \end{longtabu}
\end{toappendix}

%% Conclusion
 \section{Conclusion}\label{sec-conclusion}

We have presented the first algorithm for quantitative verification of
population protocols able to provide asymptotic bounds valid for any
number of agents. The algorithm is able to compute good bounds for
many of the protocols described in the literature.

The algorithm is based on the notion of stage graph, a concept that
can be of independent value. In particular, we think that stage graphs
can be valuable for fault localization and perhaps even automatic
repair of ill designed protocols.

An interesting question is whether our algorithm is ``weakly
complete'', meaning that for every predicate there exists a protocol
for which our algorithm can compute the exact time bound. We know that
this is the case for protocols with leaders, and conjecture that the
result extends to all protocols, but currently we do not have a proof.

Another venue for future research is the automatic computation of
lower bounds. Here, while stage graphs will certainly be useful, they
do not seem to be enough, and will have to be complemented with other
techniques.

\bibliography{ref}

\clearpage
\appendix

\end{document}